\documentclass[10pt,
superscriptaddress,
preprintnumbers,
nofootinbib,
amsmath,amssymb,
aps,
prx,
twocolumn
]{revtex4-2}
\usepackage{graphicx}
\usepackage{braket}
\usepackage{leftindex}
\usepackage{bm}
\usepackage{blkarray}
\usepackage{xcolor}
\usepackage{hyperref}

\begin{document}
\title{Measurement-based simulation of lattice gauge theory dynamics\\
with adaptive quantum circuits on a trapped-ion processor}
\author{Hiroki Sukeno}
\affiliation{Center for Quantum Information and Control, Department of Physics and Astronomy, University of New Mexico, Albuquerque, New Mexico 87106, USA}
\email{hsukeno@unm.edu}
\author{Enrico Rinaldi}
\affiliation{Quantinuum, Partnership House, Carlisle Place, London SW1P 1BX,  United Kingdom}
\email{enrico.rinaldi@quantinuum.com}
\author{Takuya Okuda}
\affiliation{Graduate School of Arts and Sciences, University of Tokyo\\
Komaba, Meguro-ku, Tokyo 153-8902, Japan}
\email{takuya@hep1.c.u-tokyo.ac.jp}
\begin{abstract}
Measurement-based quantum simulation (MBQS)---a recently proposed architecture for simulating lattice gauge theories---implements Hamiltonian dynamics by consuming a model-specific entangled resource state with adaptive mid-circuit measurements, rather than by a gate-based circuit.
The local constraints in lattice gauge theories are mirrored by the higher-form symmetries of the resource state.
Here we report, to our knowledge, the first experimental realization of MBQS of real-time dynamics in the $(2+1)$-dimensional $\mathbb{Z}_2$ gauge theory using the Quantinuum System Model H2 trapped-ion processor.
We observe coherent evolution of gauge-invariant observables on $2\times2$ and
$3\times3$ spatial lattices, consuming virtual three-dimensional cluster states
of 200 and 288 resource-state qubits that are generated from instantaneous
blocks of 48 and 54 qubits within the 56-qubit register by measurement, reset,
and re-entanglement.
The measurement record that drives the evolution simultaneously provides one-form-symmetry syndromes at no additional cost, enabling postselection that strongly suppresses observed Gauss-law violations and improves aggregate agreement with ideal Trotterized dynamics.
Our results demonstrate that MBQS is a viable, symmetry-aware architecture for simulating lattice field theories on present-day hardware.
\end{abstract}
\maketitle

\begin{figure*}[t]
	\centering
	\includegraphics[width=\linewidth]{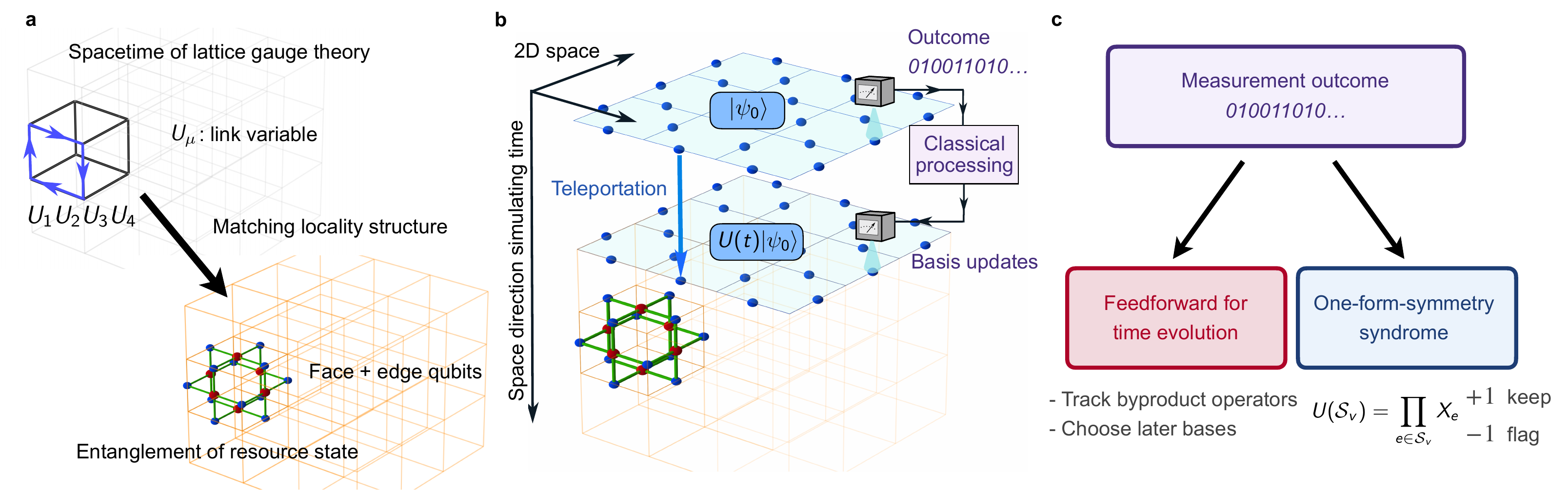}
	\caption{\textbf{Measurement-native architecture for lattice-gauge dynamics.}
	(a) A three-dimensional cluster resource state encodes the spacetime-local interaction structure of the target lattice gauge theory, most clearly exhibited by the Euclidean action given in
	Eq.~\eqref{eq:Z2-LGT-action}.
	(b) Adaptive measurements consume the resource and teleport the simulated state through the virtual time direction while applying the desired Trotterized Hamiltonian evolution.
	(c) The mid-circuit measurement record has two roles: it supplies feedforward information for Pauli byproduct tracking and later measurement bases, and it supplies one-form-symmetry syndromes for postselection.
	}
	\label{fig:architecture}
\end{figure*}

\section{Introduction}
Lattice gauge theory (LGT)~\cite{Wilson:1974sk} provides a nonperturbative formulation of quantum field theory and underlies first-principles studies of strongly interacting systems in particle and nuclear physics~\cite{Creutz:1983njd,Kronfeld:2012uk}.
Euclidean lattice methods have enabled quantitative calculations of many static and equilibrium properties~\cite{FlavourLatticeAveragingGroupFLAG:2024oxs}, but they are poorly suited to real-time dynamics, where analytic continuation and sign problems obstruct direct access to observables such as transport, scattering, and nonequilibrium response~\cite{Aarts:2026uiu}.
Quantum simulation offers a complementary route: by encoding gauge-field degrees of freedom on quantum hardware and implementing Hamiltonian time evolution directly, it can in principle probe LGT dynamics beyond the reach of classical Monte Carlo methods~\cite{Banuls:2019bmf,Bauer:2022hpo,Bauer:2023qgm}.
Experimentally, this program has progressed from few-qubit Schwinger-model dynamics~\cite{2016Natur.534..516M} to higher-dimensional gauge-theory simulations on trapped-ion, superconducting-qubit, and qudit processors~\cite{2025NatCo..16.5492M,2025arXiv250708088C,2025NatPh..21..570M}.

The highly symmetric formulation of a gauge theory is enabled by the use of an enlarged, redundant Hilbert space, but physical states are constrained to reside in the gauge-invariant subspace defined by Gauss's law~\cite{Weinberg:1995mt}.
A crucial requirement in any faithful implementation of a gauge theory is thus the preservation of gauge invariance.
Since noise and implementation errors can violate Gauss law constraints, causing the simulated state to become unphysical, detecting and suppressing this violation is a central challenge for experimental LGT simulation~\cite{2016RPPh...79a4401Z,Halimeh:2020gauge-reliability,2025arXiv250708088C}.
Mid-circuit measurements provide a natural way to extract gauge-symmetry syndromes during the computation, while feedforward or postselection can use this information to condition the subsequent dynamics or the retained ensemble, in close analogy with syndrome-based quantum error correction.
Adaptive quantum circuits are therefore a promising framework for gauge-theory quantum simulation.

Measurement-based quantum computation (MBQC) provides a natural framework for such adaptive circuits: entanglement is prepared in advance in a resource state, and computation proceeds through sequential measurements whose bases may depend on earlier outcomes~\cite{2001PhRvL..86..910B,2003PhRvA..68b2312R}.
In the standard universal setting, however, large-scale MBQC demonstrations remain experimentally challenging because the resource state must support programmable logical gates, leading to substantial overhead.
This motivates a more specialized measurement-based approach to quantum simulation, in which the resource state is tailored directly to the target physical model rather than to universal programmable computation.

Measurement-based quantum simulation (MBQS) applies the idea of MBQC in a model-specific rather than universal setting~\cite{2023ScPP...14..129S,2024PhRvR...6d3018O,2024ScPP...17..113O}.
The resource state is designed so that its entanglement connectivity follows the spacetime locality of the target field theory, allowing the desired Hamiltonian time evolution to be generated by adaptive measurements without compiling it into a universal gate set.
For local field theories, designing the resource state to follow the spacetime locality of the target model keeps the resource-state entanglement sparse.
In gauge theories, the same structure gives the resource state local symmetry generators, whose measurement outcomes can serve as syndromes for detecting errors associated with gauge-constraint violations.
More broadly, measurement-based time evolution protocols have also been developed for fermionic quantum simulation~\cite{2022PhRvR...4c2013L}.
Related measurement-based variational approaches have been demonstrated experimentally for static energy estimation, including a small $\mathbb{Z}_2$ lattice gauge theory instance~\cite{2024PhRvL.132x0601C}.
Formulations with adaptive measurements have also been proposed for Schwinger-model VQE ansatz circuits~\cite{2024PhRvD.109k4508S}.
By contrast, the present work realizes, to our knowledge, the first experimental measurement-based quantum simulation of real-time lattice gauge theory dynamics.

In this work, we turn the MBQS construction for lattice gauge theories into an experimental architecture for measurement-driven real-time gauge dynamics.
The architecture is summarized in Fig.~\ref{fig:architecture}.
We implement adaptive measurement protocols for the $(2+1)$-dimensional $\mathbb{Z}_2$ lattice gauge theory on the Quantinuum System Model H2 trapped-ion processor (H2-2)~\cite{2021Natur.592..209P,PhysRevX.13.041052}, using measurement, reset, and re-entanglement to realize virtual three-dimensional cluster states with hundreds of resource-state qubits from
a 56-qubit register.
In this architecture, the same mid-circuit measurement record both drives the Trotterized Hamiltonian evolution and provides one-form-symmetry syndromes of the resource state.
We use these syndromes for postselection and independently measure the Gauss-law generators of the output state.
The retained ensemble shows strong suppression of Gauss-law violations, while gauge-invariant observables exhibit coherent evolution and improved aggregate agreement with ideal Trotterized dynamics.

\section{Measurement-based protocol for LGT dynamics}
Discrete gauge theories provide a natural testbed for measurement-based quantum simulation of lattice gauge dynamics.
In particular, the $(2+1)$-dimensional $\mathbb{Z}_2$ gauge theory retains key structural features of gauge theories--local Gauss law constraints, nonlocal loop observables, and confinement--deconfinement phases--while using finite-dimensional link variables that can be encoded directly in qubits~\cite{RevModPhys.51.659}.
It therefore isolates the experimental problem at the center of this work: realizing real-time gauge dynamics while using syndromes for postselection and output-state measurements of the Gauss-law generators to diagnose gauge invariance.

Consider a two-dimensional square lattice on a torus of size $L_x \times L_y$, with vertex set $\Lambda_V$ satisfying $|\Lambda_V|=L_xL_y$.
The $\mathbb{Z}_2$ gauge degrees of freedom are represented by qubits placed on the links of the lattice, so that the number of LGT qubits is $N_{\rm LGT}=2L_xL_y$.
The Hamiltonian is
\begin{align}
H = - \sum_{\ell \in L} X_\ell - \lambda \sum_{p \in P} \prod_{\ell \subset p} Z_\ell \, , \label{eq:Z2LGT_Hamiltonian}
\end{align}
where $L$ and $P$ denote the sets of links and plaquettes, respectively, and $X_\ell$ and $Z_\ell$ are Pauli operators acting on link $\ell$.
The two terms are the electric and magnetic contributions to the gauge-theory Hamiltonian.
Physical states are restricted to the gauge-invariant sector by the Gauss law constraint
$G({\bf x})|\psi\rangle=|\psi\rangle$, where
$G({\bf x})=\prod_{\ell \supset {\bf x}} X_\ell$ is the product over links incident on the lattice site ${\bf x}$.
Because $[H,G({\bf x})]=0$ for all sites ${\bf x}$, ideal time evolution preserves this physical sector.

Let $E$ and $F$ denote the sets of edges and faces of the three-dimensional cubic lattice used as the MBQS resource.
We associate one resource qubit with each $e\in E$ and $f\in F$, and initialize all resource qubits in $|+\rangle$, the $+1$ eigenstate of the Pauli $X$ operator.
For any set $\mathcal Q$ of qubit labels, we write
\begin{equation}\label{eq:tensor-def}
|+\rangle^{\mathcal Q}:= \bigotimes_{q\in \mathcal Q}| +\rangle_q \,.
\end{equation}
Entanglement is generated by applying a controlled-$Z$ gate between every incident face-edge pair, where
${\rm CZ}_{i,j}=|0\rangle\langle 0|_i\otimes I_j+|1\rangle\langle 1|_i\otimes Z_j$.
The resulting three-dimensional cluster state is~\cite{2005PhRvA..71f2313R}
\begin{align}\label{eq:3D-cluster}
|\Psi\rangle =
\Big(\prod_{f\in F}\prod_{e\subset f}{\rm CZ}_{f,e}\Big)
|+\rangle^{E\cup F} \, .
\end{align}
We implement the target real-time dynamics using the symmetric second-order Trotter formula
\begin{align}
U(t) =
&\Big(
\prod_{\ell \in L }e^{ i \delta t X_\ell/2} \nonumber \\
& \times
\prod_{p \in P} e^{ i \delta t \lambda \prod_{\ell \subset p}Z_\ell}
\prod_{\ell \in L }e^{ i \delta t X_\ell/2}
\Big)^{N_t} \, ,
\label{eq:second-order}
\end{align}
where $t=N_t\delta t$.

In the measurement-based implementation~\cite{2023ScPP...14..129S}, the Trotter factors in Eq.~\eqref{eq:second-order} are generated by measuring the three-dimensional resource state rather than by applying gates directly on the LGT qubits.
Measurements of face qubits implement the magnetic and electric rotation angles, while previous measurement outcomes determine the signs of later rotation angles through classical feedforward.
This feedforward is the adaptivity of the protocol: it compensates the outcome-dependent Pauli byproduct operators produced by earlier measurements, so that each retained trajectory implements the same target Trotter evolution after classical processing.
The explicit layer-by-layer measurement pattern is given in Appendix~\ref{sec:measurement-pattern}.
For the three-dimensional LGT resource, resource layer $z$ ($z=0,\ldots,N_t-1$) is the $xy$-oriented slice of six-qubit unit cells at fixed resource coordinate $z$, bounded by the planes at $z$ and $z+1$.

The adaptive face-qubit measurements impose a causal order on part of the measurement pattern, because later measurement bases depend on earlier outcomes.
By contrast, the edge-qubit measurements are all in the fixed \(X\) basis and are not determined by any preceding measurement outcome.
They can therefore be performed in parallel, before the adaptive face-qubit measurements, while their outcomes are retained as syndrome data.
After these edge measurements, the remaining resource state is a three-dimensional toric-code state~\cite{2024PhRvX..14b1040T,2024CmPhy...7..205I}.
Equivalently, the protocol realizes LGT dynamics by adaptively measuring a topologically ordered resource state.

\section{Virtual cluster states on a trapped-ion device}
\begin{figure*}[t]
	\centering
	\includegraphics[width=\linewidth]{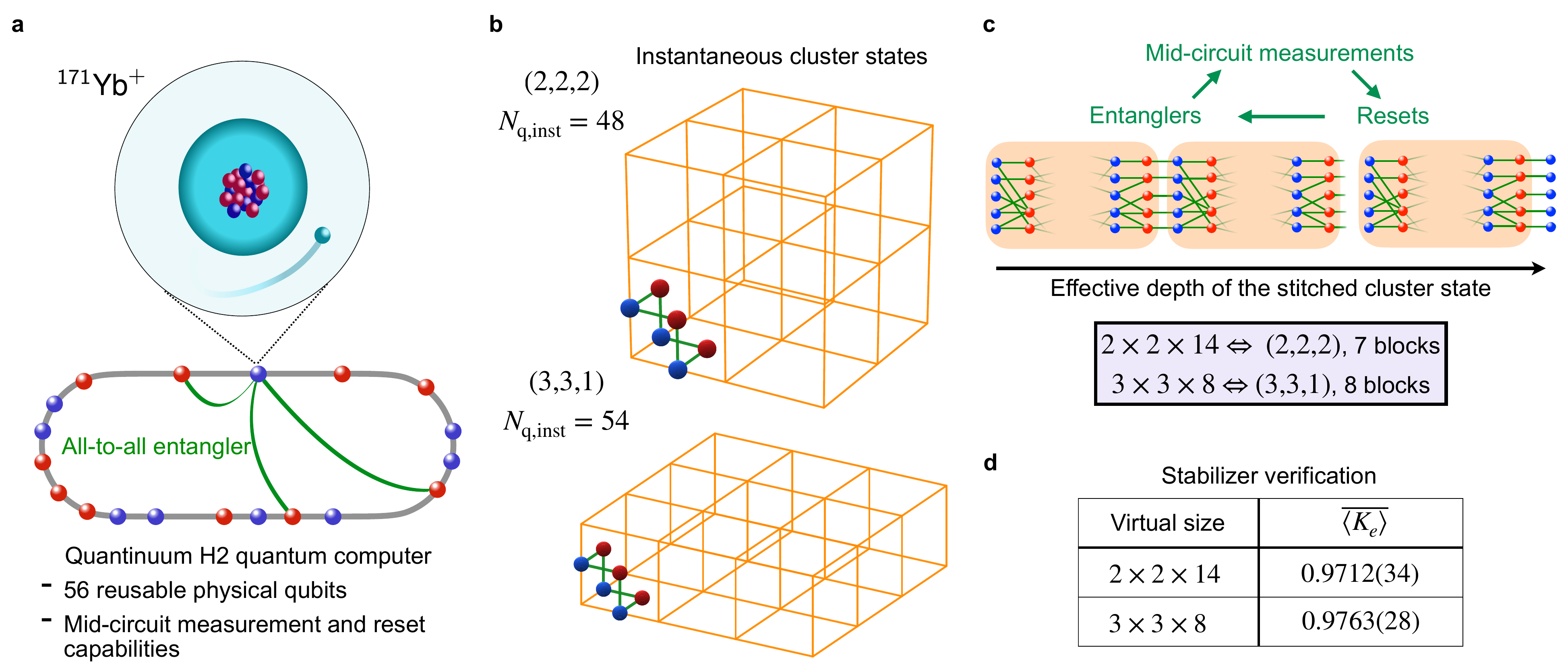}
	\caption{\textbf{Hardware realization of the virtual cluster state.}
	(a) The Quantinuum System Model H2 provides a 56-qubit trapped-ion register with mid-circuit measurement and reset capabilities.
	(b) To fit within this hardware register, the MBQS resource is decomposed into instantaneous cluster states, including $(L_x,L_y,L_z)=(2,2,2)$ with $N_{\rm q,inst}=48$ and $(3,3,1)$ with $N_{\rm q,inst}=54$.
	The figure shows a representative set of six qubits within each unit cell.
	Arranging $L_xL_yN_t$ copies of this six-qubit unit cell in an $L_x\times L_y\times N_t$ array and adding
	$2L_xL_y$ output edge qubits on the separate $xy$-plane at $z=N_t$ gives the full cluster state shown in Fig.~\ref{fig:architecture} and defined in Eq.~\eqref{eq:3D-cluster}.
	(c) Measurement, reset, and re-entanglement are repeated to stitch successive instantaneous cluster states into an extended virtual cluster state.
	(d) As a hardware benchmark of the stitched resource, edge-stabilizer measurements give $\overline{\langle K_e\rangle}=0.9763(28)$ on a virtual $3\times3\times8$ cluster state ($N_{\rm q}=450$) from $100$ shots and $0.9712(34)$ on a virtual $2\times2\times14$ cluster state ($N_{\rm q}=344$) from $106$ shots.
	Parentheses denote one standard error of the mean.}
	\label{fig:hardware-growth}
\end{figure*}
The resource-state size is a major practical challenge for measurement-based protocols.
For a cluster state of depth $N_t$, the number of virtual resource qubits is
\begin{equation}
	N_{\rm q}=2L_xL_y(3N_t+1)\,.
\end{equation}
This count includes the spacetime volume contribution as well as the boundary contribution; the boundary term uses our convention that one temporal boundary carries only the output LGT edge qubits, with no face qubits.
We address this challenge by recycling physical qubits.
When $N_{\rm q}$ exceeds the number of qubits available on the device, the protocol is implemented block by block: an instantaneous cluster state with $N_{\rm q,inst}$ qubits is prepared and measured adaptively, measured qubits are reset and re-entangled to grow the next block, and this cycle is repeated until the full virtual cluster state has been consumed (see Appendix~\ref{sec:stitching}).

The Quantinuum System Model H2 provides 56 physical qubits with mid-circuit reset and reuse~\cite{DeCross:2022kuu}.
For the decompositions used here, $L_z$ divides $N_t$, and we partition the $N_t$ resource layers into $R=N_t/L_z$ recycled instantaneous cluster-state blocks, each spanning $L_z$ consecutive resource layers and containing $N_{\rm q,inst}=6L_xL_yL_z<56$ qubits. The separate output $xy$-plane at $z=N_t$ is not part of any recycled block.
In the experiments, we use $(L_x,L_y,L_z)=(2,2,2)$ and $(3,3,1)$, corresponding to simulated lattice gauge theories with $N_{\rm LGT}=8$ and $18$ link qubits, respectively.
This hardware decomposition and the growth of the stitched virtual resource are summarized in Fig.~\ref{fig:hardware-growth}a--c.

As a first experimental benchmark, we validate the stitched MBQS resource state by measuring the edge stabilizers
\begin{equation}
	K_e = X_e \prod_{f \supset e} Z_f\,,
\end{equation}
which have eigenvalue $+1$ for the ideal cluster state in Eq.~\eqref{eq:3D-cluster}.
For virtual $3\times3\times8$ and $2\times2\times14$ cluster states constructed by stitching recycled instantaneous blocks, the all-edge averaged stabilizer values are
$\overline{\langle K_e\rangle}=0.9763(28)$ from $100$ shots and
$\overline{\langle K_e\rangle}=0.9712(34)$ from $106$ shots, respectively, as summarized in Fig.~\ref{fig:hardware-growth}d.
Fixed-$z$ resolved stabilizer data are provided in Appendix~\ref{sec:plane-resolved}.
This experiment measures only the edge-type stabilizers $K_e$.
The face-type stabilizers $K_f=X_f\prod_{e\subset f}Z_e$ are expected to behave similarly by approximate edge-face duality, although they were not measured.

\begin{figure*}
	\begin{center}
	\includegraphics[width=0.90\linewidth]{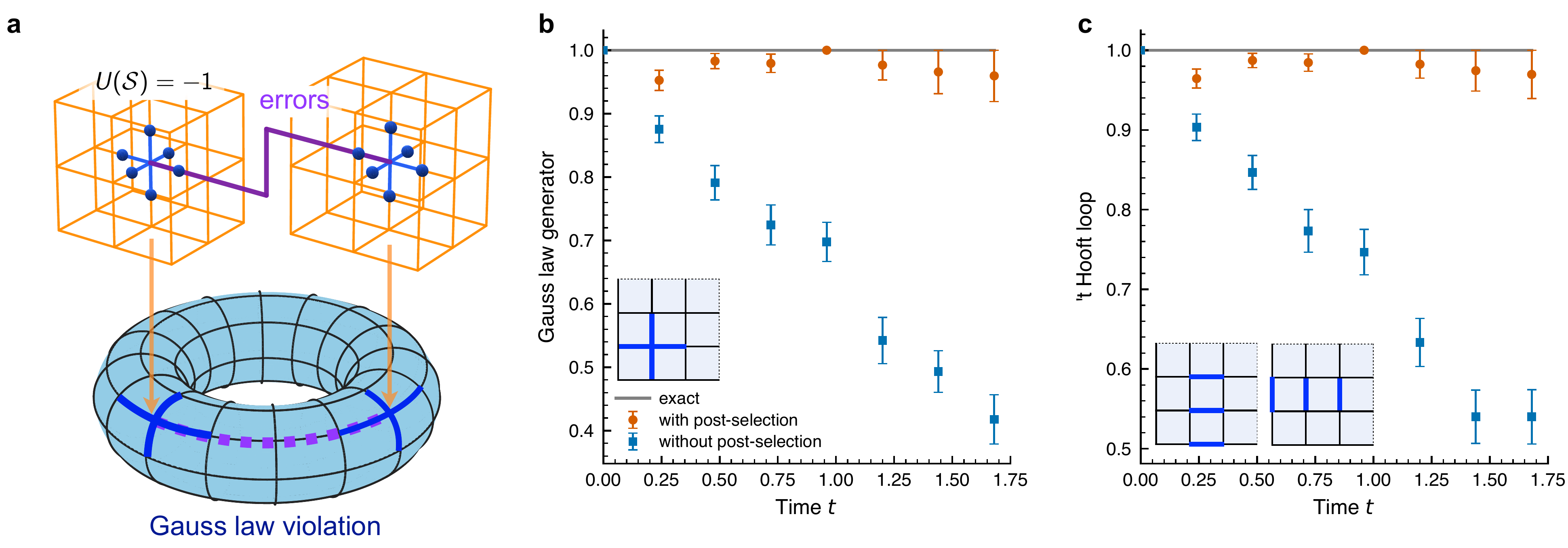}
	\end{center}
	\caption{\textbf{Syndrome-based postselection for gauge invariance.}
	(a) In the ideal 3D cluster state, the one-form-symmetry operator
	$U(\mathcal{S})=\prod_{e\in\mathcal{S}}X_e$ has eigenvalue $+1$ on any closed dual surface $\mathcal{S}$.
	For the postselection used here, $\mathcal{S}_v$ is the minimal closed dual surface around a vertex $v$, corresponding to the six edge qubits incident on that vertex in the three-dimensional resource lattice.
	Errors on edge qubits that flip the eigenvalue of $U(\mathcal{S}_v)$ are detected as resource-state syndrome violations and can lead to Gauss-law violations in the simulated two-dimensional state.
	We retain only trajectories with $U(\mathcal{S}_v)=+1$ for all checked minimal surfaces.
	(b,c) Experimental results for MBQS with $(L_x,L_y,L_z)=(3,3,1)$ instantaneous cluster states, $\lambda=2.0$, $\delta t=0.12$, and $N_{\rm shot}=100$.
	The final time $t=1.68$ is obtained with $N_{\rm q}=774$ virtual qubits and $1512$ entangling gates (each executed as one native $ZZ(\pi/2)$ gate; Appendix~\ref{app:device-info}).
	(b) Spatial average of the Gauss-law generator, $\sum_{{\bf x}\in\Lambda_V}\langle G({\bf x})\rangle/(L_xL_y)$, before and after syndrome postselection.
	(c) Expectation value of the averaged 't Hooft loop operator~\eqref{eq:avr_thooft}.
	Because these quantities are conserved by the ideal Trotterized evolution and have unit expectation value for the initial state $|+\rangle^{L}$, panels (b,c) are conserved-diagnostic tests rather than nontrivial Hamiltonian-dynamics benchmarks.}
	\label{fig:experiment_Gauss}
\end{figure*}

\section{Syndrome-based postselection for gauge invariance}
\label{sec:postselection}

The edge stabilizers $K_e$
imply a one-form symmetry condition
$U(\mathcal{S})|\Psi\rangle=|\Psi\rangle$, where
$U(\mathcal{S})=\prod_{e\in\mathcal{S}}X_e$ for a closed dual two-dimensional surface $\mathcal{S}$ in the three-dimensional resource lattice~\cite{2016PhRvB..93o5131Y,2023ScPP...14..129S}.
The eigenvalue of $U(\mathcal{S})$ serves as a syndrome for one-form symmetry violation in the resource state, rather than as a direct measurement of the Gauss-law generator $G({\bf x})$ in the simulated two-dimensional gauge theory.
Within the stochastic Pauli-error model of Ref.~\cite{2023ScPP...14..129S} and under the boundary idealization detailed in Appendix~\ref{sec:one-form-symmetry-syndrome}, trivial values of all local one-form-symmetry syndromes imply a gauge-invariant output state.
We use these syndromes for postselection and independently test this implication through the Gauss-law measurements of the output state in Fig.~\ref{fig:experiment_Gauss}b.

In each execution of the MBQS protocol, we maintain a classical postselection flag $b_{\rm ps}$, initialized to $0$.
Let $m_\mu(i,j,k)\in\{0,1\}$ denote the $X$-basis measurement outcome on the resource-state edge in direction $\mu=x,y,z$ originating at the vertex $(i,j,k)$.
For every spatial vertex and every resource layer $k=0,\ldots,N_t-1$, we evaluate $U(\mathcal S_{i,j,k})=(-1)^{s_{i,j,k}}$, where
\begin{align}
s_{i,j,k} = {} & m_z(i,j,k) + m_z(i,j,k-1) \nonumber\\
&+ m_x(i,j,k)+m_x(i-1,j,k) \nonumber \\
&+m_y(i,j,k)+m_y(i,j-1,k) \mod 2 \,.
\end{align}
The spatial indices are periodic.
At the initial temporal boundary we set $m_z(i,j,-1)=0$, so the $k=0$ syndrome contains five measured outcomes and one fixed $+1$ boundary factor.
No additional syndrome is evaluated at the output boundary $k=N_t$.
Thus, the total number of checked syndromes is $N_{\rm syn}=L_xL_yN_t$.
Whenever $s_{i,j,k}=1$, the flag is set to $b_{\rm ps}=1$.
In postprocessing, we retain only trajectories with $b_{\rm ps}=0$.
The Gauss-law observables of the retained simulated states are then evaluated independently, as shown below.

We test the Hamiltonian-evolution protocol by diagnosing Gauss-law preservation in the simulated two-dimensional LGT state.
We measure the spatial average of the Gauss-law generator,
$\sum_{{\bf x}\in\Lambda_V}\langle G({\bf x})\rangle/(L_xL_y)$.
This quantity equals one for an ideal gauge-invariant trajectory, while deviations indicate Gauss-law violations in the simulated state.
The results before and after resource-state syndrome postselection are shown in Fig.~\ref{fig:experiment_Gauss}b.

We further characterize the simulated LGT dynamics using three gauge-invariant observables.
The first is the averaged 't Hooft loop over the two noncontractible cycles of the torus (shown in Fig.~\ref{fig:experiment_Gauss}c),
\begin{align}
\frac{1}{L_x +L_y }\Bigg(\sum_{i=1}^{L_x}  \prod_{\ell \in V_x(i)} X_{\ell}
+
\sum_{j=1}^{L_y}  \prod_{\ell \in V_y(j)} X_{\ell}\Bigg)
\, ,
\label{eq:avr_thooft}
\end{align}
where $V_x(i)$ ($V_y(j)$) is the set of $x$-directional ($y$-directional) links whose coordinates satisfy $x \in [i,i+1]$ ($y \in [j,j+1]$), with $i$ ($j$) understood modulo $L_x$ ($L_y$).
Notice that $V_x(i)$ ($V_y(j)$) runs over a $y$-directional ($x$-directional) straight line.
The second is the averaged Wilson loop operator
\begin{align}
\overline{W_{m\times n}}:=
\frac{1}{L_x L_y}\sum_{{\bf x} \in \Lambda_V}  \langle W_{m\times n}({\bf x})\rangle
\, ,
\label{eq:avr_plaquette}
\end{align}
where $W_{m\times n}({\bf x})$ is the product of $Z$ operators on links in the $m\times n$ rectangle whose bottom-left corner is at ${\bf x}$.
We experimentally measure $\overline{W_{1\times 1}}$, $(\overline{W_{1\times 2}} + \overline{W_{2\times 1}})/2$, and $\overline{W_{2\times 2}}$.
The third is the average electric energy:
\begin{align}
\frac{1}{N_{\rm LGT}} \sum_{\ell \in L} \langle X_\ell \rangle \, .
\label{eq:avr_electric}
\end{align}

\begin{figure*}
	\centering
\includegraphics[width=\linewidth]{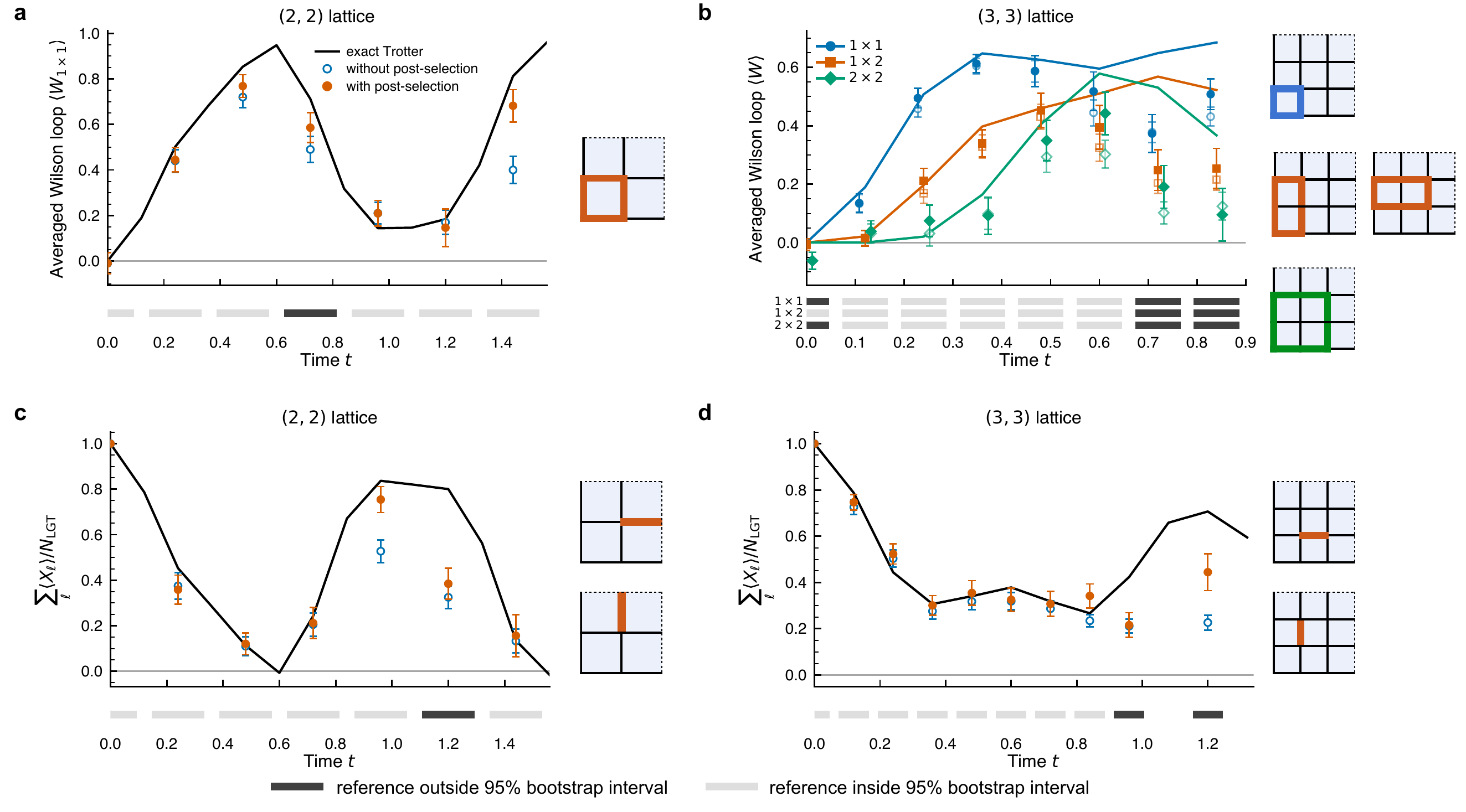}
 \caption{
    \textbf{Coherent dynamics of gauge-invariant observables in $(2+1)$D LGT.}
	Experimental MBQS results for $\lambda=2.0$, $\delta t=0.12$, and $N_{\rm shot}=100$.
	Data are shown before and after postselection on the one-form-symmetry syndrome, together with the ideal reference curves; error bars denote one standard error of the mean.
	The strip beneath each panel reports, for every measured time, the outcome of a consistency test performed independently of the plotted error bars: a dark cell marks a data point at which the exact second-order Trotterized reference lies outside the central $95\%$ bootstrap percentile interval of the postselected estimate, obtained by resampling the $N_{\rm shot}$ shot records of each time and measurement basis with replacement ($N_{\rm boot}=10^5$); a light gray cell marks consistency at this level.
	(a) Experiment using $(L_x,L_y,L_z)=(2,2,2)$ instantaneous cluster states with $N_{\rm q,inst}=48$.
	The expectation value of the spatial average of the minimal plaquette operators is plotted.
	(b) Experiment using $(L_x,L_y,L_z)=(3,3,1)$ instantaneous cluster states with $N_{\rm q,inst}=54$.
	The expectation values of the spatial average of the Wilson loop operators with sizes $1\times 1$, $1\times2$ (including $2\times 1$), and $2\times 2$ are plotted.
	(c) Similar data for the expectation values of the spatial average of the electric energy density operator $\sum_{\ell \in L} \langle X_\ell\rangle/N_{\rm LGT}$ for $(L_x,L_y,L_z)=(2,2,2)$.
	(d) Similar data with $(L_x,L_y,L_z)=(3,3,1)$.
    }
	\label{fig:experiment_result}
\end{figure*}

The Gauss-law generator and the 't Hooft loop are both products of Pauli-$X$ operators and are obtained from the same final Pauli-$X$ readout basis, while the averaged Wilson loop operator requires the corresponding Pauli-$Z$ readout.
The Gauss-law average and the averaged 't Hooft loop are conserved diagnostics for the ideal dynamics: for the initial link state $|+\rangle^{L}$, with $L$ denoting the set of spatial links, their exact Trotterized reference curves remain at unity.
The averaged Wilson loop operator and the electric energy operator, by contrast, give the nontrivial dynamics benchmark in Fig.~\ref{fig:experiment_result}.  We evaluate these observables before and after postselection on the one-form-symmetry syndrome and compare them with the corresponding exact reference curves.

The experimentally measured  dynamics in Fig.~\ref{fig:experiment_result} reproduce the early-time behavior of the exact Trotterized evolution and extend to the onset of visible deviations at the largest depths.
As the evolution depth increases, the unpostselected data show a gradual drift from the exact curve, while postselection on trajectories with no detected resource-state one-form-symmetry violation improves the aggregate agreement with the exact evolution.
See Appendix~\ref{sec:postselection-quantitative}, where we quantitatively assess the effectiveness of postselection.

For each curve in Fig.~\ref{fig:experiment_result} we consider the running maximum deviation of the postselected estimates $\widehat O^{\rm ps}(t_i)$ from the exact Trotterized reference $r(t_i)$ over the displayed times $t_i$, bounded at $68\%$ bootstrap confidence:
\begin{align}
\Delta_{68}(T)
= q_{68}\!\left[\,\max_{t_i \le T}\,\bigl|\widehat O^{\rm ps}(t_i) - r(t_i)\bigr|\,\right] ,
\label{eq:coherence-window}
\end{align}
where $q_{68}$ denotes the 68th percentile of the shot-level bootstrap distribution (Appendix~\ref{sec:postselection-quantitative}); with at least $68\%$ bootstrap confidence, every deviation up to time $T$ lies below $\Delta_{68}(T)$.
Choosing a threshold $\Delta_{68}(T)<0.20$, all observables of the $(2,2,2)$ experiment remain below threshold up to $t=0.96$, where the resource state consists of $N_{\rm q}=200$ virtual qubits and $384$ entangling gates (each executed as one native $ZZ(\pi/2)$ gate; Appendix~\ref{app:device-info}). 
All observables of the $(3,3,1)$ experiment remain below threshold up to $t=0.60$, where the resource state consists of $N_{\rm q}=288$ virtual qubits and $540$ entangling gates.
These windows are unchanged at the looser threshold $0.25$.

Although the objective differs from previous experimental MBQC
demonstrations~\cite{2005Natur.434..169W,2013PhRvL.111u0501L,2025NatCo..16..106R,2025PhRvL.135p0801G,2026NatSR..16..689K,2026NatPh..22..430J},
the virtual cluster states consumed to demonstrate the coherent real-time dynamics here ($N_{\rm q}=200$ and $288$; within the $\Delta_{68}(T)<0.20$ threshold) are, to our
knowledge, among the largest qubit-based cluster states consumed in an MBQC
experiment: the largest previously reported are a simultaneously prepared
95-qubit resource~\cite{2026NatPh..22..430J} and 52- and 56-vertex patterns in
which mid-circuit measurement and reset extend the pattern beyond the physical
register~\cite{2025PhRvL.135p0801G,2026NatSR..16..689K}, as they do here.

\section{Experimental scale and resource structure }
\label{sec:scale-resource}

The distinction between physical register size and virtual resource size is illustrated in Fig.~\ref{fig:hardware-growth}a--c.
Using mid-circuit measurement, reset, and re-entanglement, the 56-qubit H2 register realizes virtual resources containing hundreds of cluster-state qubits.
For the coherent dynamics within the $\Delta_{68}(T)<0.20$ threshold in Fig.~\ref{fig:experiment_result}, the $(2,2,2)$ instantaneous cluster states use $N_{\rm q,inst}=48$ qubits and generate a virtual resource with $N_{\rm q}=200$ qubits and $384$ entangling gates,
while the $(3,3,1)$ instantaneous cluster states use $N_{\rm q,inst}=54$ qubits and generate a virtual resource 
with $N_{\rm q}=288$ qubits and $540$ entangling gates.
The stitched resource is benchmarked with the virtual $3\times3\times8$ and $2\times2\times14$ cluster states of Fig.~\ref{fig:hardware-growth}d, which contain $N_{\rm q}=450$ and $344$ resource-state qubits.
For the syndrome-postselection experiment in Fig.~\ref{fig:experiment_Gauss}, the same $(3,3,1)$ instantaneous cluster-state geometry realizes a larger virtual resource with $N_{\rm q}=774$ qubits and $1512$ entangling gates.

Throughout, entangling-gate counts refer to the CZ gates of the circuit-level protocol; for the resource geometry used here, $N_{\rm CZ}=12L_xL_yN_t=2N_{\rm q}-4L_xL_y$, growing linearly with the simulated spacetime volume.
As compiled and executed on the hardware, every entangling gate is realized as exactly one native arbitrary-angle two-qubit gate $ZZ(\theta)=e^{-i(\theta/2)\,Z\otimes Z}$ at $\theta=\pi/2$ --- CZ being equivalent to $ZZ(\pi/2)$ up to single-qubit $Z$ rotations and a global phase --- so the quoted counts are simultaneously the native two-qubit-gate counts of the executed circuits (Appendix~\ref{app:device-info}).

Here, the relevant resource comparison is between the present model-specific MBQS construction and a universal-MBQC implementation of the same Trotter circuit, rather than the direct gate-based implementation itself.
By encoding the spacetime-local interaction structure directly in the resource-state connectivity, MBQS avoids the need to compile the corresponding gate-based circuit into universal-MBQC measurement patterns~\cite{2003PhRvA..68b2312R}.
Because each spacetime cell contributes only a constant number of CZ gates, the resulting resource state remains sparsely entangled.
Together with qubit recycling, this resource structure helps make the protocol experimentally feasible on current all-to-all trapped-ion processors.

\section{Limitations and outlook}
\label{sec:limit-outlook}

The present one-form-symmetry syndrome protocol uses postselection rather than deterministic correction.
For the syndrome-postselection data in Fig.~\ref{fig:experiment_Gauss}b,c, we acquired $N_{\rm shot}=100$ shots at each plotted time point.
At $n=0,2,4,6,8,10,12,14$ Trotter steps, corresponding to $t=0,0.24,0.48,0.72,0.96,1.20,1.44,1.68$, the counts of accepted shots were
$N_{\rm acc}=100,75,52,43,34,19,13,11$.
Thus, the final plotted point uses an acceptance rate of $0.11$, illustrating the sampling cost of using postselection as the demonstrated protection mechanism.

A useful way to quantify this cost is to regard postselection as the event that all checked one-form-symmetry syndromes are trivial.
If each of the $N_{\rm syn}$ checks has an effective probability $q_{\rm eff}$ of being nontrivial, then $P_{\rm acc}\simeq (1-q_{\rm eff})^{N_{\rm syn}}\simeq\exp(-q_{\rm eff}N_{\rm syn})$.
For the $(3,3,1)$ data at $N_t=14$, $N_{\rm syn}=126$, and the observed $P_{\rm acc}=0.11$ gives $q_{\rm eff}=1-P_{\rm acc}^{1/N_{\rm syn}}\simeq 0.0174$.
Because neighboring checks share measurement outcomes and device errors may be correlated, $q_{\rm eff}$ is only a phenomenological summary, not a microscopic per-check error probability.
At the final point, retaining $100$ accepted trajectories would therefore require about $9\times 10^2$ raw shots.
This sampling overhead motivates moving from postselection toward active online syndrome decoding.

Such syndrome-based correction has already been formulated for the MBQS setting in Ref.~\cite{2023ScPP...14..129S}.
In that construction, nontrivial one-form-symmetry syndromes identify endpoints of open edge-$Z$ error chains in the resource state: a recovery chain, for example one obtained by minimum-weight matching~\cite{Edmonds1965PathsTA}, as in related surface-code decoding contexts~\cite{2002JMP....43.4452D}, can then be incorporated into the classical feedforward instead of rejecting the shot.
Realizing correction-based protection would require a hardware control stack capable of incorporating decoding results into the adaptive measurement and feedforward loop during the computation, rather than using the syndrome record only after completing the run to reject completed trajectories.
This capability is beyond what is implemented in the present H2 experiment.
Developing such fast decoding and adaptive feedforward capabilities, which are also central to scalable MBQC more broadly, is therefore an important direction for making syndrome protection scalable beyond postselection.

Relative to universal MBQC, the advantage of the present construction is architectural, as discussed in Sec.~\ref{sec:scale-resource}; relative to direct gate-based simulation, any advantage of MBQS is platform dependent.
In the idealized resource accounting of Ref.~\cite{2023ScPP...14..129S}, both schemes have sequential runtimes linear in the number of Trotter steps; once the one-time resource-state preparation term assumed in that model becomes subleading, MBQS is favored when single-qubit measurements are fast relative to CZ gates, at the cost of a resource whose qubit count grows with the evolution depth.
In the qubit-recycled architecture used here, the resource is instead prepared and consumed block by block, so an end-to-end advantage additionally requires repeated re-entanglement, measurement, reset, adaptive feedforward, and syndrome processing to be sufficiently fast and high fidelity, with postselection overhead included in comparisons at fixed target accuracy.
While the present experiment does not establish such an overall advantage over the gate-based implementation, it demonstrates the measurement-based and recycling architecture on which any such advantage would be built.

The present experiment realizes one member of a broader measurement-based framework for gauge theories.
At the theoretical level, MBQS already extends from the $\mathbb{Z}_2$ model studied here to higher-dimensional Wegner models~\cite{Wegner} with higher-form Abelian gauge fields and to finite cyclic gauge groups $\mathbb{Z}_N$ using qudit generalized cluster states~\cite{2023ScPP...14..129S}.
Abelian gauge theories with matter can also be incorporated while retaining sparse resource-state connectivity~\cite{2024PhRvR...6d3018O}.
These results identify concrete next experimental targets, including $\mathbb{Z}_N$ link variables on qudit-capable processors~\cite{2022NatPh..18.1053R,2023NatCo..14.2242H}, building on recent qudit-based lattice-gauge-theory simulation experiments~\cite{2025NatPh..21..570M}.
A more open theoretical direction is to combine adaptive measurement-based protocols with finite-dimensional formulations of non-Abelian gauge dynamics, such as quantum-group and string-net approaches to $\mathrm{SU}(N)$ gauge theories~\cite{2023PhRvL.131q1902Z,2023JHEP...09..126H,2023JHEP...09..123H,2026arXiv260113530H}.
Beyond trapped ions, neutral-atom platforms with mid-circuit measurement, reset, and reconfigurable entangling operations may provide a complementary route to implementing MBQS~\cite{2026PhRvL.136p0601Y,2025arXiv250720009S,2026arXiv260315561L}.

\section*{Acknowledgments}
We thank Dan Mills and Kevin Hemery for feedback on the manuscript.
The work of H.S. was supported by National Science Foundation Grant No. PHY-2116246.
The research of T.O. was supported in part by JST PRESTO Grant Number JPMJPR23F3.
\clearpage
\appendix

\section{Resource state and one-form symmetry}
Our general design principle for the entangled resource state is that its entanglement connectivity follows the spacetime locality of the simulated field theory~\cite{2023ScPP...14..129S,2024PhRvR...6d3018O}.
Consider a three-dimensional cubic lattice with boundaries at $z=0$ and $z=N_t$.
We take the $x$ and $y$ directions to be periodic.
Let $E$ and $F$ be the sets of edges and faces of the lattice, respectively.
Let $\sigma_e\in\{\pm 1\}$ be the classical spin variable on edge $e$.
The Euclidean field theory action for the $\mathbb{Z}_2$ LGT is given by
\begin{align}\label{eq:Z2-LGT-action}
S_{\rm E}= - K \sum_{f\in F} \prod_{ e \subset f} \sigma_e \, ,
\end{align}
where $K$ is the coupling constant.
The Euclidean path integral is related to the quantum Hamiltonian~\eqref{eq:Z2LGT_Hamiltonian} via the transfer matrix formalism~\cite{RevModPhys.51.659}.

In this MBQS protocol, the $z$ direction of the three-dimensional resource is used as the simulated time direction.
The resource state is the three-dimensional cluster state defined in Eq.~\eqref{eq:3D-cluster}, whose graph reflects the face-edge incidence structure appearing in $S_{\rm E}$, see also Ref.~\cite{2007PhRvL..98k7207D}.
Equivalently, resource qubits are placed on faces and edges of the cubic lattice, and CZ gates entangle incident face-edge pairs.
Because the resource graph is tied to the spacetime lattice of the gauge theory, the cluster state has nontrivial higher-form symmetries~\cite{2016PhRvB..93o5131Y,2015JHEP...02..172G}.
As discussed in Appendix~\ref{sec:one-form-symmetry-syndrome}, the corresponding symmetry eigenvalues provide syndrome information useful for detecting gauge-constraint violations during the measurement-based simulation.

\section{Measurement pattern for the MBQS protocol}

\label{sec:measurement-pattern}
Here we spell out the measurement pattern used to implement the time-evolution operator $U(t)$ in Eq.~\eqref{eq:second-order}.
For $N_t=0$, no measurement pattern is needed beyond state preparation.
For $N_t\geq 1$, the symmetric Trotter product can be rewritten as
\begin{align}
\quad U(t)
&=
\prod_{\ell \in L }e^{ i \delta t X_\ell/2}
\prod_{p \in P} e^{ i \delta t \lambda \prod_{\ell \subset p}Z_\ell} \nonumber \\
&\quad \times
\Big(
\prod_{\ell \in L }e^{ i \delta t X_\ell}
\prod_{p \in P} e^{ i \delta t \lambda \prod_{\ell \subset p}Z_\ell}
\Big)^{N_t-1}
\nonumber\\
&\quad\times\prod_{\ell \in L }
e^{ i \delta t X_\ell/2}
\, .
\label{eq:second-order-reshuffled}
\end{align}
We choose the initial LGT state to be $|+\rangle^{L}$, which is translationally invariant and satisfies the Gauss law constraint.
Since this state is a simultaneous $+1$ eigenstate of all $X_\ell$, the rightmost electric half step in Eq.~\eqref{eq:second-order-reshuffled} contributes only an overall phase and can be omitted in the measurement pattern.
After omitting this rightmost half step, the measurement pattern realizes, in chronological order, a magnetic plaquette step, alternating electric and magnetic steps, and a final electric half step.
We denote these elementary unitaries by
$v_\ell(\xi_2)=\exp(i\xi_2 X_\ell)$, with $\xi_2=\delta t$ or $\delta t/2$, and
$u_p(\xi_1)=\exp(i\xi_1\prod_{\ell\subset p}Z_\ell)$, with $\xi_1=\lambda\delta t$.

Generically, measurement outcomes generate Pauli byproduct operators on the encoded LGT qubits.
Throughout the measurement sequence, we keep the classical record, $b_X(\ell),b_Z(\ell)\in\{0,1\}$, of byproducts
\begin{equation}
\Sigma(\{b_X(\ell),b_Z(\ell)\}) =
\prod_{\ell\in L} X_\ell^{b_X(\ell)} Z_\ell^{b_Z(\ell)} \,.
\label{eq:byproduct}
\end{equation}
We encode each measurement eigenvalue as $(-1)^m$, where $m\in\{0,1\}$, and all updates of the byproduct bits below are modulo two.
The bits are initialized as $b_X(\ell)=b_Z(\ell)=0$.
For a horizontal face $f$, let $p(f)\in P$ denote the associated spatial LGT plaquette.
When the face qubit is measured in the eigenbasis of $\mathcal{O}_A(\xi_1(f))$ defined below, with outcome bit $m_f^A$, the adaptive angle and byproduct-bit update are
\begin{align}
&\xi_1(f) = (-1)^{\sum_{\ell\subset p(f)}b_X(\ell)} \lambda \delta t \,, \nonumber \\
& b_Z(\ell)  \leftarrow b_Z(\ell) + m_f^A    \mod 2\qquad (\ell \subset p(f)) \,. \label{eq:bZ-update1}
\end{align}
Thus, because $b_X(\ell)=0$ initially, the first magnetic step has $\xi_1(f)=\lambda\delta t$ for every boundary face $f$.
An $X$-basis measurement of a horizontal edge qubit associated with link $\ell$, with outcome bit $m_\ell^X$, updates the Pauli-byproduct record according to
\begin{equation}
b_Z(\ell)\leftarrow b_Z(\ell)+ m_\ell^X \mod 2 \,.\label{eq:bZ-update2}
\end{equation}
By contrast, the fixed $X$-basis measurements of vertical edge qubits used to compute the one-form-symmetry syndromes do not update the Pauli-byproduct record.
For a vertical face associated with link $\ell$ and measured in the eigenbasis of $\mathcal{O}_B(\xi_2)$ defined below, with outcome bit $m_\ell^B$, the adaptive angle and byproduct-bit update are
\begin{align}
\xi_2(\ell) &= (-1)^{b_Z(\ell)}\delta t \,, \nonumber \\
b_X(\ell) & \leftarrow b_X(\ell) + m_\ell^B \mod 2 \,,\label{eq:bX-update}
\end{align}
with $\delta t$ replaced by $\delta t/2$ in the final electric half step.
After the final resource layer has been measured, the recorded Pauli byproduct $\prod_{\ell \in L} X_\ell^{b_X(\ell)}Z_\ell^{b_Z(\ell)}$ is removed by applying the corresponding Pauli corrections to the output state.
The ordering of these corrections can change only a physically irrelevant global phase.

At the initial boundary
$xy$-plane
at $z=0$, the input LGT state is supported on the boundary edge qubits.
We first measure each boundary face qubit in the eigenbasis of
$\mathcal{O}_A(\xi_1)=e^{-i\xi_1 X}Ze^{i\xi_1 X}$.
For a boundary face $f$, this measurement implements the magnetic plaquette unitary $u_{p(f)}(\xi_1(f))=\exp(i\xi_1(f)\prod_{\ell\subset p(f)}Z_\ell)$ on the adjacent boundary edge qubits~\cite{2023ScPP...14..129S}.
Taking $\xi_1(f)=\lambda\delta t$ for all boundary faces gives the first magnetic step $\prod_{p\in P} u_p(\lambda\delta t)$.
At this stage, we get byproduct operators~\eqref{eq:byproduct} with~$b_X(\ell)=0$.
We then measure the boundary edge qubits in the $X$-basis; this teleports the encoded LGT state to the next
$xy$-plane of edge qubits in the resource state and modifies byproduct operators~\eqref{eq:byproduct} by updating~$b_Z(\ell)$.

Within resource layer $z$, the qubits in the slab between the bounding $xy$-planes at $z$ and $z+1$ implement the next electric step and teleport the state from the former plane to the latter.
The vertical edge qubits in this slab are measured in the fixed $X$ basis.
These measurements pass the LGT state and the byproduct operators unchanged, and the measurement outcomes are used to compute the resource-state one-form-symmetry syndromes $U(\mathcal{S}_v)$.
If $U(\mathcal{S}_v)=-1$ for any checked vertex~$v$, a classical flag $b_{\rm ps}$, initialized to 0, is set to 1.
This flag is used for postselection; the syndromes are not, by themselves, direct measurements of the simulated $G({\bf x})$.

After these fixed-basis edge measurements, the face qubits in the same slab are measured in the eigenbasis of
$\mathcal{O}_B(\xi_2)=e^{-i\xi_2 Z}Xe^{i\xi_2 Z}$.
Because earlier measurements have generated outcome-dependent byproducts, the sign of $\xi_2(\ell)$ is chosen by classical feedforward from the current $Z$-byproduct bit $b_Z(\ell)$, according to the rules (\ref{eq:bZ-update1}) and (\ref{eq:bZ-update2}); its magnitude is $\delta t$ for a full electric step and $\delta t/2$ for the final half step.
These measurements implement the electric evolution $\prod_{\ell} v_\ell(\xi_2)$ and modify Pauli byproducts, so that the classical bits $\{b_X(\ell),b_Z(\ell)\}$ are updated from the measurement outcomes.

This slab-by-slab procedure is then repeated through the final resource layer.
For each subsequent magnetic step, the qubit on each relevant horizontal face $f$ is measured in the eigenbasis of $\mathcal{O}_A(\xi_1(f))$. The sign of $\xi_1(f)$ is chosen according to Eq.~\eqref{eq:bZ-update1} from the parity $\sum_{\ell\subset p(f)}b_X(\ell)$, where the current $b_X$ record has been updated according to Eq.~\eqref{eq:bX-update}.
Subsequent measurements follow the same pattern as above and update rules of byproduct operators as we described above.
Together with the electric-step measurements described above, this adaptive layer-by-layer pattern implements the Trotterized evolution $U(t)$ with $t=N_t\delta t$, up to the tracked Pauli byproducts.
Throughout the protocol, the edge-qubit measurements remain fixed $X$-basis measurements; the adaptive choices enter only through the signs of the face-qubit measurement angles.

\section{MBQS with qubit recycling}
\label{sec:stitching}

We formalize the qubit-recycling construction used in the experiment as an equivalence between two implementations of the same measurement-based computation.
In Model A, the full graph-state resource is prepared at once and then consumed by adaptive measurements.
In Model B, only one instantaneous block of the graph is held in the hardware register at a time.
After the relevant qubits in a block are measured, they are reset to $|+\rangle$ and re-entangled to the next block.
Generalizing the recycling argument for the 2D cluster state~\cite{2003PhRvA..68b2312R}, we show below that, for block-decomposable graphs with causal measurement adaptivity, Model B produces the same conditional output state as Model A for each measurement record.

\begin{figure}[b]
	\includegraphics[width=\linewidth]{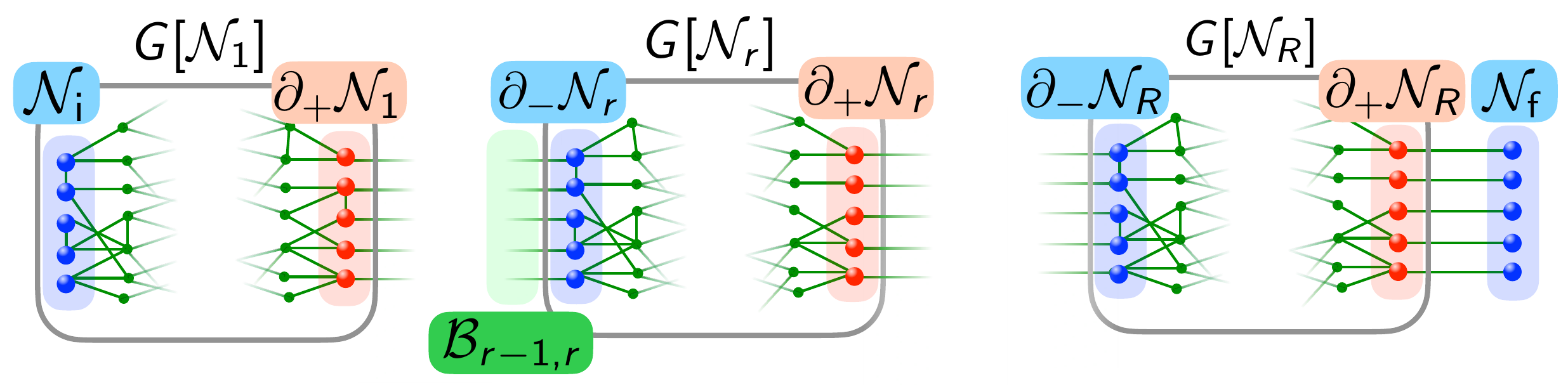}
	\caption{\textbf{Block decomposition of the resource graph.}
	The graph $G=(\mathcal{N},\mathcal{B})$ is divided into ordered recycled blocks $G[\mathcal{N}_r]$ and a separate output block $\mathcal{N}_{\rm f}$, with incoming and outgoing boundary registers $\partial_-\mathcal{N}_r$ and $\partial_+\mathcal{N}_r$.
	Inter-block bonds $\mathcal{B}_{r-1,r}$, including the final matching $\mathcal{B}_{R,R+1}$ from $\partial_+\mathcal{N}_R$ to $\mathcal{N}_{\rm f}$, connect neighboring blocks, allowing the stitched Model-B implementation to reproduce the full-resource Model-A computation.}
	\label{fig:graph}
\end{figure}

Let $G=(\mathcal{N},\mathcal{B})$ be the graph of the resource state, with node set $\mathcal{N}$ and bond set $\mathcal{B}$. We decompose the nodes into $R$ ordered recycled blocks together with a separate output block,
\begin{align}
\mathcal{N}=\Big(\bigsqcup_{r=1}^R\mathcal{N}_r\Big)\sqcup\mathcal{N}_{\rm f},
\end{align}
and choose the input logical register $\mathcal{N}_{\rm i}\subset\mathcal{N}_1$; the output logical register is the separate block $\mathcal{N}_{\rm f}$, which is not identical to the recycled blocks $\mathcal{N}_r$. Throughout we use the shorthand $\mathcal{N}_{R+1}\equiv\mathcal{N}_{\rm f}$, so that the last recycled block $\mathcal{N}_R$ and the output block are treated on the same footing in the interface formulas below. For notational uniformity, set
\begin{align}
\partial_-\mathcal{N}_1:=\mathcal{N}_{\rm i}.
\end{align}
For the interfaces, define
\begin{align}
\partial_+\mathcal{N}_r
=
\{u\in\mathcal{N}_r:\exists v\in\mathcal{N}_{r+1}\ {\rm with}\ (u,v)\in\mathcal{B}\} \nonumber\\
\quad (r=1,\ldots,R),
\end{align}
and
\begin{align}
\partial_-\mathcal{N}_r
=
\{u\in\mathcal{N}_r:\exists v\in\mathcal{N}_{r-1}\ {\rm with}\ (v,u)\in\mathcal{B}\} \nonumber\\
\quad (r=2,\ldots,R),
\end{align}
together with $\partial_-\mathcal{N}_{\rm f}=\mathcal{N}_{\rm f}$, since every node of the output block is matched to $\mathcal{N}_R$. The inter-block bond sets are
\begin{align}
\mathcal{B}_{r,r+1}
&:=
\{(u,v)\in\mathcal{B}:u\in\partial_+\mathcal{N}_r,\ v\in\partial_-\mathcal{N}_{r+1}\}
\nonumber\\
&\quad (r=1,\ldots,R)\,,
\end{align}
where $\mathcal{B}_{R,R+1}$ connects $\mathcal{N}_R$ to the output block $\mathcal{N}_{\rm f}$.
We assume that
\begin{itemize}
\item[(i)] each recycled block $\mathcal{N}_r$ ($r=1,\ldots,R$) induces a connected subgraph $G[\mathcal{N}_r]$ with fixed size $|\mathcal{N}_r|=N_{\rm q,inst}$, while the separate output block $\mathcal{N}_{\rm f}$ may differ in size and carries no internal bonds,
\item[(ii)] every bond lies either within a single block or between two neighboring blocks, 
\item[(iii)] the incoming and outgoing boundary registers of each recycled block are disjoint, $\partial_-\mathcal{N}_r\cap\partial_+\mathcal{N}_r=\emptyset$,
\item[(iv)] all boundary registers have the same size, $|\partial_-\mathcal{N}_r|=|\partial_+\mathcal{N}_r|=|\mathcal{N}_{\rm i}|=|\mathcal{N}_{\rm f}|$,
\item[(v)] each $\mathcal{B}_{r,r+1}$ ($r=1,\ldots,R$) forms a perfect matching between the corresponding boundary registers.
\end{itemize}
Thus each recycled block has an incoming boundary $\partial_-\mathcal{N}_r$ and an outgoing boundary $\partial_+\mathcal{N}_r$, and the output block has only an incoming boundary $\partial_-\mathcal{N}_{\rm f}=\mathcal{N}_{\rm f}$. Conditions (ii), (iv), and (v) define how the logical register is passed from one block to the next, and finally to $\mathcal{N}_{\rm f}$, while conditions (i) and (iii) allow each recycled block to be implemented on a fixed-size instantaneous hardware register. See Fig.~\ref{fig:graph} for an illustration.

In Model A, the entire graph-state resource on $G$ is prepared before the adaptive measurement pattern is applied.
The input state $|\psi_0\rangle$ is placed on $\mathcal{N}_{\rm i}$, all other resource qubits are initialized in $|+\rangle$, and the graph-state entangler
\begin{align}\label{eq:graph-CZ}
U^{\mathcal{N}}_{\rm CZ}:=\prod_{(u,v)\in\mathcal{B}}{\rm CZ}_{u,v}
\end{align}
is applied.
We assume that the measurement pattern is compatible with the block order: when a qubit in $\mathcal{N}_r$ is measured, its basis may depend only on outcomes from qubits that have already been measured in $\mathcal{N}_{\le r}:=\bigsqcup_{q=1}^r\mathcal{N}_q$. For a fixed measurement record $\vec m$ and the corresponding adaptive angles $\vec\theta$, the conditional output state on $\mathcal{N}_{\rm f}$ is
\begin{align}
&
|\varphi^{(A)} (\vec{m}, \vec{\theta})\rangle
\nonumber \\
&=  \frac{1}{\sqrt{Z_{\vec{m},\vec{\theta}}}}\Big( \bigotimes_{n \in
\mathcal{N}\setminus\mathcal{N}_{\rm f}
}
\langle m_n,\theta_n|  \Big)\cdot
U^\mathcal{N}_{\rm CZ} \Big( |\psi_0\rangle_{\mathcal{N}_{\rm i}} \otimes |+\rangle^{\mathcal{N} \setminus \mathcal{N}_{\rm i}}\Big) \, ,
\end{align}
where $\langle m_n,\theta_n|$ denotes projection onto outcome $m_n$ in the measurement basis specified by $\theta_n$, and $Z_{\vec m,\vec\theta}$ is the corresponding normalization factor. Here
$\vec m=\{m_n\}_{n\in\mathcal{N}\setminus\mathcal{N}_{\rm f}}$ and
$\vec\theta=\{\theta_n\}_{n\in\mathcal{N}\setminus\mathcal{N}_{\rm f}}$.

Model B realizes the recycled blocks sequentially on a single $N_{\rm q,inst}$-qubit hardware register. To write this process as an operator product, we identify each recycled block $\mathcal{N}_r$ ($r=1,\ldots,R$) with a fixed reference register, which we take to be $\mathcal{N}_1$. More explicitly, for each $r$ we choose a bijection $\iota_r:\mathcal{N}_1\rightarrow\mathcal{N}_r$ that maps the incoming boundary of the reference register to $\partial_-\mathcal{N}_r$ and the outgoing boundary of the reference register to $\partial_+\mathcal{N}_r$.
With the convention specified above, this includes the endpoint identification $\partial_-\mathcal{N}_1=\mathcal{N}_{\rm i}$.
The output block $\mathcal{N}_{\rm f}$, with $|\mathcal{N}_{\rm f}|=|\partial_+\mathcal{N}_R|$ by condition (iv), is realized on the outgoing-boundary sub-register left in $|+\rangle$ by the final stitch; it is not one of the recycled blocks and is not measured.
The bijection need not preserve the internal bond structure of the induced subgraphs $G[\mathcal{N}_r]$, so the intra-block entangling operation may depend on $r$. In what follows, operators labelled by $\mathcal{N}_r$ are understood, through this identification, as operators acting on the same $N_{\rm q,inst}$-qubit register.

We now define the elementary maps used in Model B. 
Let, for $r=1,\ldots,R$,
\begin{align}
\mathcal{B}_r:=\{(u,v)\in\mathcal{B}:u,v\in\mathcal{N}_r\},
\quad
U^{\mathcal{N}_r}_{\rm CZ}:=\prod_{(u,v)\in\mathcal{B}_r}{\rm CZ}_{u,v}
\end{align}
be the intra-block entangler; the output block carries no such entangler, since $\mathcal{N}_{\rm f}$ has no internal bonds. For $r=1,\ldots,R$, define the measurement-reset map
\begin{align}
M_r:=\bigotimes_{n\in\mathcal{N}_r\setminus\partial_+\mathcal{N}_r}
|+\rangle\langle m_n,\theta_n|,
\end{align}
which measures the consumed qubits of block $r$ and resets them to $|+\rangle$. The output block $\mathcal{N}_{\rm f}$ carries no measurement map, since it holds the conditional output state. For each interface let
\begin{align}
S_{r,r+1}:=
\prod_{n\in\partial_+\mathcal{N}_r}
\left[
\Big(|+\rangle\langle m_n,\theta_n|\Big)_n\,{\rm CZ}_{n,n_+}
\right] \nonumber \\ 
\quad (r=1,\ldots,R),
\end{align}
where $n_+\in\partial_-\mathcal{N}_{r+1}$ is the vertex matched to $n$ by $\mathcal{B}_{r,r+1}$; for $r=R$ these target vertices lie in $\mathcal{N}_{\rm f}$. The recycled-block maps are then
\begin{align}
W_1 &= M_1U^{\mathcal{N}_1}_{\rm CZ}, \\
W_r &= M_rU^{\mathcal{N}_r}_{\rm CZ}S_{r-1,r}\quad(1<r\le R),
\end{align}
and the final output map is the single stitch
\begin{align}
W_{\rm f} = S_{R,R+1},
\end{align}
which entangles the outgoing boundary of $\mathcal{N}_R$ with $\mathcal{N}_{\rm f}$ and measures and resets $\partial_+\mathcal{N}_R$, leaving $\mathcal{N}_{\rm f}$ unmeasured.
Thus the Model-B conditional output state is
\begin{align}
&\quad|\varphi^{(B)}(\vec m,\vec\theta)\rangle
\nonumber\\
&=
\frac{1}{\sqrt{Z_{\vec m,\vec\theta}}}
W_{\rm f}W_R W_{R-1}\cdots W_1
\left(
|\psi_0\rangle_{\mathcal{N}_{\rm i}}
\otimes
|+\rangle^{\mathcal{N}_1\setminus\mathcal{N}_{\rm i}}
\right).
\end{align}
Expanding the product $W_{\rm f}W_R W_{R-1}\cdots W_1$ shows that it contains exactly the same operations as Model A, but ordered block by block.
The intra-block entanglers $U^{\mathcal{N}_r}_{\rm CZ}$ generate all CZ gates inside each recycled block, while the stitching maps $S_{r,r+1}$ ($r=1,\ldots,R$) generate the CZ gates across neighboring blocks, including the final matching to $\mathcal{N}_{\rm f}$, and then measure and reset the outgoing boundary register.
Since condition (ii) excludes bonds beyond neighboring blocks, condition (v) gives a one-to-one matching across each interface, and $\mathcal{N}_{\rm f}$ has no internal bonds, these operations reproduce the full graph-state entangler $U^{\mathcal{N}}_{\rm CZ}$ after relabeling by the maps $\iota_r$.
The block-causal adaptivity assumption below Eq.~\eqref{eq:graph-CZ} ensures that the measurement angle $\theta_n$ used in Model B is the same function of the previous measurement outcomes as in Model A. Therefore, for every measurement record $\vec m$ and corresponding adaptive angles $\vec\theta$,
\begin{align}\label{eq:equivalence-AB}
|\varphi^{(A)}(\vec m,\vec\theta)\rangle
=
|\varphi^{(B)}(\vec m,\vec\theta)\rangle .
\end{align}
Thus the full-resource MBQS computation can be implemented with a fixed-size hardware register by repeated intra-block entangling operations, inter-block stitching operations, measurements, and resets.

The resource graphs used for the $(2+1)$D $\mathbb{Z}_2$ LGT experiments satisfy the assumptions above.
For both real-time evolution and stabilizer verification, the measurement bases are compatible with the block order: bases for qubits in block $\mathcal{N}_r$ depend only on outcomes already available from earlier measured qubits in $\mathcal{N}_{\le r}$.
Concretely, the full $(L_x,L_y,N_t)$ cluster state has $N_{\rm q}=|\mathcal{N}|=2L_xL_y(3N_t+1)$ qubits. It decomposes into $R=N_t/L_z$ identical instantaneous $(L_x,L_y,L_z)$ cluster-state blocks, each with $N_{\rm q,inst}=|\mathcal{N}_r|=6L_xL_yL_z$ qubits so that $\sum_{r=1}^R|\mathcal{N}_r|=R\,|\mathcal{N}_r|=6L_xL_yN_t$, together with the separate output block $\mathcal{N}_{\rm f}$ of $|\mathcal{N}_{\rm f}|=2L_xL_y$ qubits; these satisfy $R\,|\mathcal{N}_r|+|\mathcal{N}_{\rm f}|=N_{\rm q}$ and the graph conditions (i)--(v).
For $r=1,\ldots,R$, the recycled block $\mathcal{N}_r$ consists of the resource layers $z=(r-1)L_z,\ldots,rL_z-1$, while $\mathcal{N}_{\rm f}$ is the separate output edge register on the bounding $xy$-plane at $z=N_t$.
The equivalence of Models A and B therefore applies to the experimental recycling protocol.
In particular, for stabilizer verification, the joint distribution of measurement outcomes obtained by sequentially applying the maps $W_{\rm f}W_R\cdots W_1$ is identical to the distribution obtained by preparing the full cluster state and measuring the corresponding qubits in $\mathcal{N}\setminus\mathcal{N}_{\rm f}$.

\section{One-form-symmetry syndrome and postselection}

\label{sec:one-form-symmetry-syndrome}
The Euclidean action~\eqref{eq:Z2-LGT-action} is gauge invariant under the local $\mathbb{Z}_2$ transformation that flips all edge variables incident on an arbitrary vertex $v$.
In the MBQS resource-state description, this gauge structure is represented by closed-surface symmetry operators of the ideal cluster state~\eqref{eq:3D-cluster}.
For a closed dual surface $\mathcal{S}$ in the three-dimensional resource lattice, the corresponding operator is
\begin{align}
U(\mathcal{S})=\prod_{e\in\mathcal{S}}X_e,
\qquad
U(\mathcal{S})|\Psi\rangle=|\Psi\rangle .
\end{align}
For the local surfaces used below, $\mathcal{S}=\mathcal{S}_v$ is the minimal closed dual surface enclosing a vertex $v$, equivalently the set of six resource-lattice edges incident on $v$.
These closed-surface operators are the resource-state one-form symmetries used in the postselection scheme below.

We use the one-form-symmetry syndromes introduced in Ref.~\cite{2023ScPP...14..129S}, but apply them here by postselection rather than by an explicit recovery operation.
To describe the error-detection mechanism, we use the standard Pauli-error discretization viewpoint of quantum error correction~\cite{NielsenChuang} and model noisy MBQS evolution by a 3D cluster state corrupted by stochastic Pauli noise,
$|\Psi_{\rm corrupt}\rangle=(\prod_{e,f} X_e^{r_e}Z_e^{t_e}  X_f^{u_f} Z_f^{w_f} )|\Psi\rangle$, with $r_e,t_e,u_f, w_f\in\{0,1\}$ and ideal measurements.
For a local closed surface $\mathcal{S}_v$, the operator $U(\mathcal{S}_v)$ is sensitive to the Pauli-$Z$ error component on the edge qubits: a nontrivial value $U(\mathcal{S}_v)=-1$ indicates an odd parity of such errors on the edges in $\mathcal{S}_v$.
Equivalently, these nontrivial values diagnose endpoints of open $Z$-error strings in the resource lattice.
Postselection keeps only trajectories with the trivial syndrome, $U(\mathcal{S}_v)=+1$ for all checked vertices $v$.
As an idealization, let us assume $t_e=0$ for all the edges that are incident on the boundary plane where the output state is induced.
In those retained trajectories, the remaining $Z$-error component has no detected endpoints and is therefore restricted to endpoint-free strings,
\begin{align}
&U(\mathcal{S}_v)=+1  \text{ for all checked }v \nonumber \\
&\Rightarrow
|\Psi_{\rm corrupt}\rangle = \Big(\prod_{e,f} X_e^{r_e} X_f^{u_f} Z_f^{w_f} \Big)Z(\gamma)|\Psi\rangle \, ,
\end{align}
where $\gamma$ is a sum of closed curves, including the trivial identity operator.
For independent edge-qubit $Z$ errors with rate $p$, each undetected closed error loop of length $\ell$ occurs with probability of order $p^\ell$.

The above Pauli errors propagate to the output state during measurement-induced teleportation.
The operator $Z(\gamma)$ is teleported to the boundary plane, and it becomes a product of $Z$ operators acting on projected closed curves on the boundary gauge-theory qubits, which is gauge invariant. It can also be shown that the other types of stochastic Pauli errors---$X$-type edge errors and $X/Z$-type errors on bulk face qubits---do not lead to violation of the gauge constraint~\cite{2023ScPP...14..129S}.
Only near-boundary edge-$Z$ errors, which we assumed absent as an idealization, contribute to the gauge-constraint violation of the simulated state.
This occurs with probability $1-(1-p)^{3L_xL_y}\simeq 3L_xL_yp$.

\section{Device information}
\label{app:device-info}

We performed the hardware experiments reported here on Quantinuum's H2-2 trapped-ion quantum computer. The device follows the Quantinuum System Model H2 design: ${}^{171}\mathrm{Yb}^{+}$ ions encode qubits in ground-state hyperfine levels and are transported in a racetrack quantum charge-coupled-device architecture~\cite{Wineland:1997mg,Kielpinski:2002wbd,PhysRevX.13.041052}.
In the representative component benchmarks listed below, the largest reported component infidelities are the two-qubit-gate error rate, $8.3(5)\times10^{-4}$, and SPAM error rates, $6.7(9)\times10^{-4}$ for $|0\rangle$ and $1.2(1)\times10^{-3}$ for $|1\rangle$.

Our H2-2 runs were executed during March 2026--June 2026. Table~\ref{tab:quantinuum_metrics} lists representative component-level performance metrics reported by Quantinuum for the device. These values are taken from Quantinuum's H2-2 hardware-specification data set dated 2025-08-28, as reported in the performance-validation documentation~\cite{QuantinuumPerformance2026}.
We use these component-level metrics to indicate the representative H2-2 hardware error scale; they are not run-by-run calibrations for the March--June 2026 jobs.

In the H2-2 data-taking runs reported in this paper, we did not enable dynamical decoupling or Quantinuum's leakage-detection gadgets, although both features are available on H2 devices.
Test runs with dynamical decoupling gave changes smaller than our statistical uncertainties.
In the representative H2-2 benchmarks, the reported qubit-leakage rates are lower than the listed two-qubit-gate and SPAM error rates. This indicates that leakage was comparatively small in those benchmarks, although run-level calibration records were not available to determine its contribution during our experiments.

All circuits were constructed in \texttt{pytket}~\cite{Sivarajah:2020lfo} using Hadamard, CZ, and CNOT gates, single-qubit rotations, mid-circuit measurement and reset, and classically conditioned single-qubit gates, and were compiled with the default compilation pass of the Quantinuum Nexus platform at optimization level~2, targeting the H2-2 native gate set $\{U_{1q}(\theta,\phi), \, R_z(\lambda),\,ZZ(\theta)\}$ with $ZZ(\theta)=e^{-i(\theta/2)\,Z\otimes Z}$.
In the compiled circuits executed on hardware, every CZ and CNOT gate is realized as exactly one $ZZ(\pi/2)$ gate together with single-qubit rotations; the optimizer neither merges nor eliminates two-qubit gates for these circuits, no other two-qubit operations appear, and all dependence on the Trotter angles is carried by single-qubit rotations.
The entangling-gate counts quoted in the main text therefore coincide with the native two-qubit-gate counts of the executed circuits: for MBQS, $12L_xL_y$ native $ZZ(\pi/2)$ gates per Trotter step ($576$ for the $(2,2,2)$ run at $N_t=12$, 
$540$ for the $(3,3,1)$ run at $N_t=5$
and $1512$ for the $(3,3,1)$ postselection run at $N_t=14$); for the gate-based comparison, $10L_xL_y$ per step ($560$ for $(L_x,L_y)=(2,2)$ at $N_t=14$ and $540$ for $(3,3)$ at $N_t=6$).  We verified these counts directly on the compiled circuits submitted for execution.

\begin{table*}[t]
\centering
\begin{tabular}{lcc}
\hline
Metric & Infidelity & Uncertainty \\
\hline
1-Qubit Gate Error &
$2.8\times10^{-5}$ &
$3.6\times10^{-6}$ \\
1-Qubit Gate Leakage Error &
$6.1\times10^{-6}$ &
$1.3\times10^{-6}$ \\
2-Qubit Gate Error &
$8.3\times10^{-4}$ &
$4.8\times10^{-5}$ \\
2-Qubit Gate Leakage Error &
$1.9\times10^{-4}$ &
$1.7\times10^{-5}$ \\
Memory Error per Depth-1 Circuit Time &
$1.2\times10^{-4}$ &
$2.0\times10^{-5}$ \\
Measurement Crosstalk Error &
$2.2\times10^{-5}$ &
$5.3\times10^{-7}$ \\
SPAM Error for $|0\rangle$ &
$6.7\times10^{-4}$ &
$8.7\times10^{-5}$ \\
SPAM Error for $|1\rangle$ &
$1.2\times10^{-3}$ &
$1.1\times10^{-4}$ \\
\hline
\end{tabular}
\caption{\textbf{Quantinuum-reported H2-2 component error rates.} Representative error rates and associated uncertainties are taken from Quantinuum's H2-2 performance-validation data set dated 2025-08-28~\cite{QuantinuumPerformance2026}.}
\label{tab:quantinuum_metrics}
\end{table*}

\section{Estimating observables and uncertainties}
\label{sec:obs-unc}

We consider observables that are diagonal in the measurement basis,
\begin{equation}
\mathcal{O}=\sum_{\bm{s}} c_{\bm{s}}\prod_{i=1}^n Z_i^{s_i},
\qquad
\bm{s}=(s_1,\ldots,s_n)\in\{0,1\}^n,
\end{equation}
where the $n$ qubits encode the gauge-theory degrees of freedom. Observables related to this form by local basis changes, such as the spatial average of the stabilizers $K_e$ in the main text, are treated analogously.

Suppose that all $n$ qubits are measured $M$ times in the $Z$ basis, producing bit strings $\bm{b}^{(a)}=(b_1^{(a)},\ldots,b_n^{(a)})$ for $a=1,\ldots,M$. For each shot, define
\begin{equation}
\mathcal{O}_a
:=\sum_{\bm{s}} c_{\bm{s}}\prod_{i=1}^n (-1)^{s_i b_i^{(a)}} .
\end{equation}
We estimate the expectation value and its squared statistical uncertainty by
\begin{align}
\langle\mathcal{O}\rangle_{\mathrm{est}}
&=\overline{\mathcal{O}}
:=\frac{1}{M}\sum_{a=1}^M\mathcal{O}_a\,,
\nonumber\\
s_\mathcal{O}^2
&:=\frac{1}{M(M-1)}\sum_{a=1}^M
\left(\mathcal{O}_a-\overline{\mathcal{O}}\right)^2 .
\end{align}
By the Born rule, $\mathbb{E}[\mathcal{O}_a]=\mathrm{tr}(\rho\mathcal{O})\equiv\langle\mathcal{O}\rangle$, and therefore $\mathbb{E}[\langle\mathcal{O}\rangle_{\mathrm{est}}]=\langle\mathcal{O}\rangle$. For independent shots, $s_\mathcal{O}^2$ is the usual unbiased estimator of $\mathrm{Var}(\langle\mathcal{O}\rangle_{\mathrm{est}})$.

\section{
Layer-resolved
 stabilizer verification of stitched resource states}

\label{sec:plane-resolved}
\begin{figure*}
	\begin{center}
	\includegraphics[width=\linewidth]{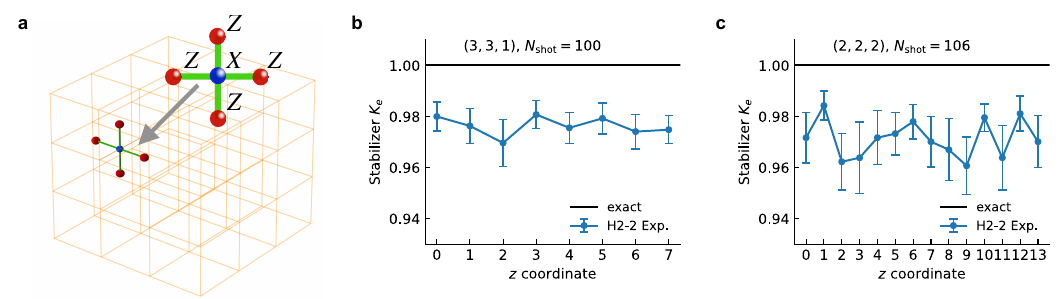}
	\end{center}
	\caption{\textbf{Stabilizers of the 3D cluster state.}
	(a) The 3D cluster state is a simultaneous $+1$ eigenstate of the stabilizers $K_e$ (depicted above) and $K_f$.
	(b)
	Layer-resolved edge-stabilizer expectation values for a virtual $3\times3\times8$ cluster state constructed by stitching eight recycled $(3,3,1)$ instantaneous blocks on the H2-2 quantum processor.
	The resource-layer coordinate $z$ runs from $0$ to $7$ and labels the eight layers of unit cells, rather than the $N_t+1$ bounding $xy$-planes.
	At each $z$, the plotted value is averaged over the $3L_xL_y$ edge stabilizers assigned to that layer: $L_xL_y$ stabilizers for edges in each of the three lattice orientations.
	The stabilizer benchmark therefore contains no separate output-boundary point at $z=N_t$.
	(c)
	Analogous results for a virtual $2\times2\times14$ cluster state constructed from seven recycled $(2,2,2)$ instantaneous blocks, each contributing two successive resource layers, with $z=0,\ldots,13$.
	}
	\label{fig:experiment_stab}
\end{figure*}

Figure~\ref{fig:hardware-growth}d summarizes the edge-stabilizer verification of the stitched resource states, giving
$\overline{\langle K_e\rangle}=0.9763(28)$ for the virtual $3\times3\times8$ cluster state and
$\overline{\langle K_e\rangle}=0.9712(34)$ for the virtual $2\times2\times14$ cluster state. Here we provide the corresponding
 resource-layer-resolved
  data.
Figure~\ref{fig:experiment_stab} shows the $3\times3\times8$ case, constructed from eight recycled $(3,3,1)$ instantaneous blocks, and the $2\times2\times14$ case, constructed from seven recycled $(2,2,2)$ instantaneous blocks.

\section{Gate-based simulation of (2+1)D $\mathbb{Z}_2$ LGT}
For comparison with MBQS, we also implement a gate-based circuit for the second-order Trotter evolution of the $(2+1)$D $\mathbb{Z}_2$ LGT on the Quantinuum H2-2 processor.
We consider an $L_x\times L_y$ square lattice with periodic boundary conditions.
The gauge degrees of freedom are encoded in one qubit on each link $\ell\in L$, and the Gauss law generator $G({\bf x})$ is defined at each lattice site ${\bf x}$.
The Hamiltonian is Eq.~\eqref{eq:Z2LGT_Hamiltonian}, and the ideal second-order Trotter step is Eq.~\eqref{eq:second-order}.
In the gate-based implementation, we use the reshuffled ordering analogous to Eq.~\eqref{eq:second-order-reshuffled} and insert projective measurements of the local Gauss law generators after each magnetic plaquette step.
Writing
$E_\alpha=\prod_{\ell\in L}e^{i\alpha X_\ell}$,
$B=\prod_{p\in P}e^{i\delta t\lambda\prod_{\ell\subset p}Z_\ell}$, and
$P_n=\prod_{{\bf x}}\{1+(-1)^{g_n({\bf x})}G({\bf x})\}/2$,
where the product over ${\bf x}$ runs over lattice sites, the implemented evolution is
\begin{align}
&|\psi_{\rm gate}(t)\rangle\nonumber \\
&=
E_{\delta t/2}
P_{N_t}B
E_{\delta t}
P_{N_t-1}B
\cdots
E_{\delta t}
P_1B
E_{\delta t/2}
|\psi_{\rm gate}(0)\rangle ,
\label{eq:second-order-Gauss}
\end{align}
where $t=N_t\delta t$, the factors act on $|\psi_{\rm gate}(0)\rangle$ from right to left, and the middle sequence is omitted when $N_t=1$.
Here $g_n({\bf x})\in\{0,1\}$ is the measured Gauss law syndrome at site ${\bf x}$ during Trotter step $n$, corresponding to the eigenvalue $(-1)^{g_n({\bf x})}$ of $G({\bf x})$.
In the implementation described below, we
use the gauge-invariant initial state $|\psi_{\rm gate}(0)\rangle=|+\rangle^L$.
In the absence of noise, all Trotter factors commute with every $G({\bf x})$, so all syndrome outcomes satisfy $g_n({\bf x})=0$; nonzero syndromes therefore signal Gauss law violations caused by device noise.

\begin{figure*}
	\centering
\includegraphics[width=\linewidth]{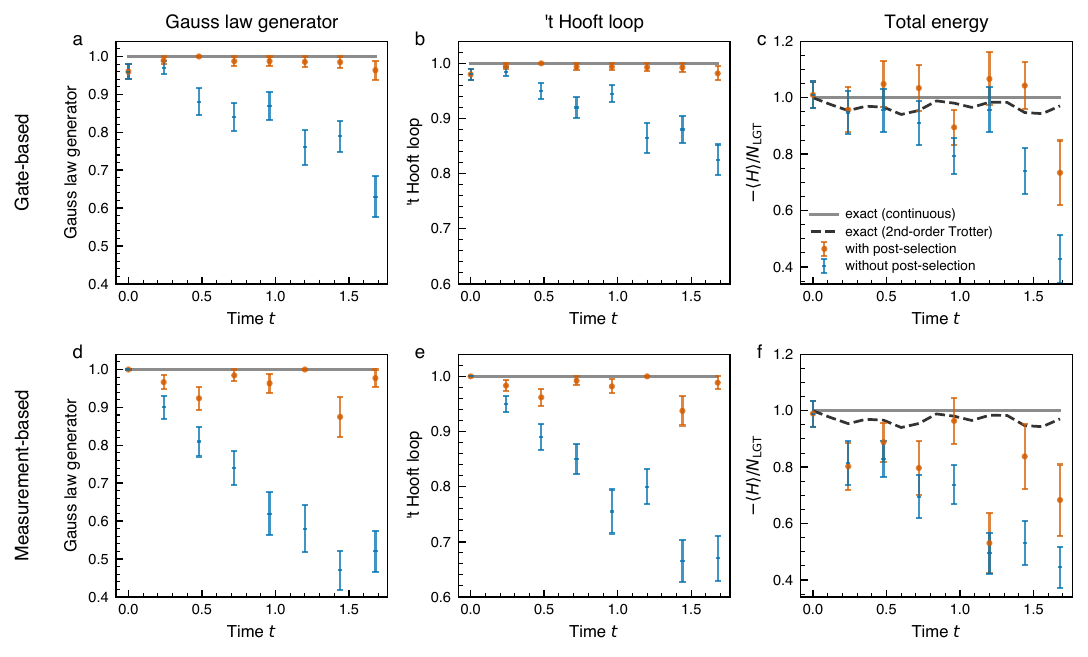}
	\caption{\textbf{Comparison between the gate-based simulation and MBQS (1).} The experiment was done for the $(L_x,L_y)=(2,2)$ spatial lattice with $\lambda = 2.0$, $\delta t= 0.12$, and $N_{\rm shot}=100$.
	{\bf Top panels (a-c)}, results from the gate-based quantum simulation.
	{\bf Lower panels (d-f)}, results from the measurement-based quantum simulation with $(L_x,L_y,L_z)=(2,2,2)$.
	(a,d) The Gauss law generator averaged over vertices of the 2D square lattice. (b,e) The 't Hooft operator over non-contractible loops.
	The expectation value of the averaged 't Hooft operator~\eqref{eq:avr_thooft} is plotted.
	(c,f) The rescaled expectation value $- \langle H \rangle/N_{\rm LGT}$ of the Hamiltonian~\eqref{eq:Z2LGT_Hamiltonian} is plotted.
	The black dashed curve is the exact second-order Trotterized evolution, the reference used in Table~\ref{tab:postselection-rmse}; the solid gray curve is the continuous-time evolution.
	The MBQS energy curve in panel (f) combines independently acquired $X$- and $Z$-basis data through $N_t=14$.
	The final component measurements at $t=1.68$, which are not displayed in panels (a) and (c) of Fig.~\ref{fig:experiment_result}, use $N_{\rm q}=344$ virtual resource qubits and 672 entangling gates.
	}
	\label{fig:GB_result_22}
\end{figure*}

\begin{figure*}
	\centering
\includegraphics[width=\linewidth]{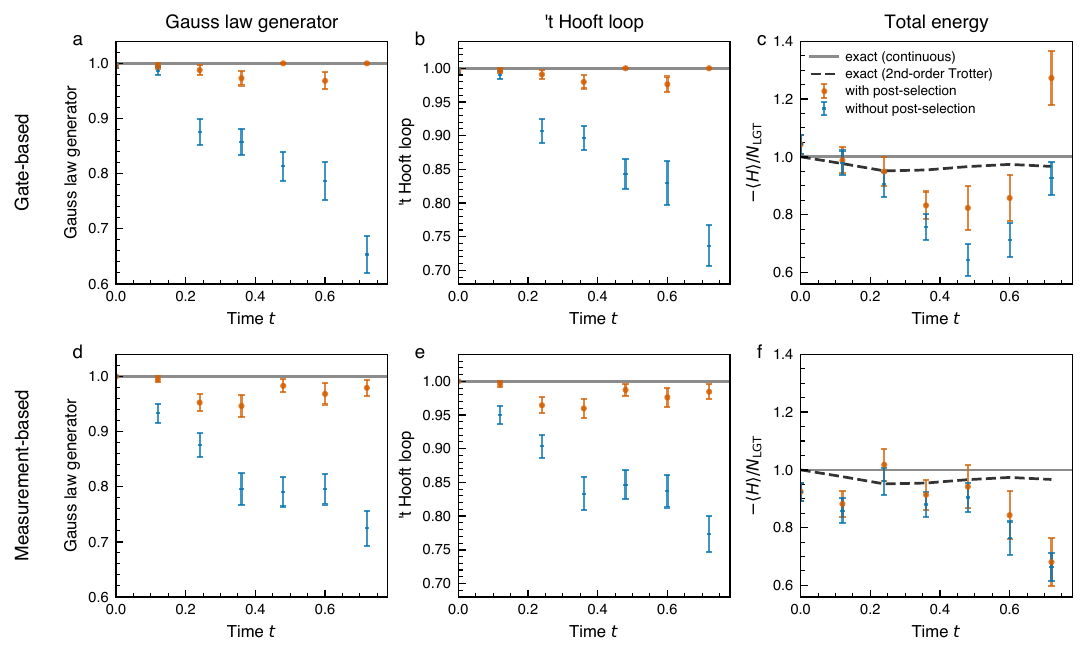}
	\caption{\textbf{Comparison between the gate-based simulation and MBQS (2).} The experiment was done for the $(L_x,L_y)=(3,3)$ spatial lattice with $\lambda = 2.0$, $\delta t= 0.12$, and $N_{\rm shot}=100$.
	{\bf Top panels (a-c)}, results from the gate-based quantum simulation.
	{\bf Lower panels (d-f)}, results from the measurement-based quantum simulation with $(L_x,L_y,L_z)=(3,3,1)$.
	(a,d) The Gauss law generator averaged over vertices of the 2D square lattice. (b,e) The 't Hooft operator over non-contractible loops.
	The expectation value of the averaged 't Hooft operator~\eqref{eq:avr_thooft} is plotted.
	(c,f) The rescaled expectation value $- \langle H \rangle/N_{\rm LGT}$ of the Hamiltonian~\eqref{eq:Z2LGT_Hamiltonian} is plotted.
	The black dashed curve is the exact second-order Trotterized evolution, the reference used in Table~\ref{tab:postselection-rmse}; the solid gray curve is the continuous-time evolution.
	}
	\label{fig:GB_result_33}
\end{figure*}

We realize the ingredients in Eq.~\eqref{eq:second-order-Gauss} using single-qubit rotations, Hadamard gates, CZ and CNOT gates, and mid-circuit measurements on the Quantinuum H2-2 processor, with each CZ and CNOT compiled to one native $ZZ(\pi/2)$ gate as described in Appendix~\ref{app:device-info}.
The initial state $|\psi_{\rm gate}(0)\rangle=|+\rangle^L$ is prepared by initializing all link qubits in $|0\rangle$ and applying Hadamard gates.
The projective Gauss law measurement and the four-body magnetic plaquette rotation are decomposed as follows:
\begin{itemize}
\item To measure $\{1+(-1)^{g_n({\bf x})}G({\bf x})\}/2$, label the four links incident on site ${\bf x}$ by $1,2,3,4$, and label the ancilla qubit for this site by $0$.
The ancilla is initialized in $|+\rangle_0$, coupled to the four incident link qubits by CNOT gates $\prod_{k=1}^4{\rm CX}_{0,k}$, and measured in the $X$ basis.
The measurement outcome gives the syndrome bit $g_n({\bf x})\in\{0,1\}$.
\item To implement $e^{i\delta t\lambda\prod_{\ell\subset p}Z_\ell}$, label the four links around plaquette $p$ by $1,2,3,4$.
We apply a Hadamard gate to qubit $1$, apply CZ gates $\prod_{k=2}^4{\rm CZ}_{k,1}$, apply the single-qubit rotation $e^{i\delta t\lambda X_1}$, then undo the same CZ gates and the Hadamard gate on qubit $1$.
\end{itemize}
Per Trotter step, the gate-based circuit thus uses $6$ CZ gates per plaquette and $4$ CNOT gates per Gauss-law ancilla, i.e.\ $10L_xL_y$ two-qubit gates per step, each executed as one native $ZZ(\pi/2)$ gate.
We run the gate-based circuit that implements Eq.~\eqref{eq:second-order-Gauss} on the Quantinuum H2-2 device without error correction or error mitigation beyond Gauss law syndrome postselection.
Specifically, we retain only shots for which $g_n({\bf x})=0$ for every lattice site ${\bf x}$ and every Trotter step $n$, and discard all shots with at least one detected Gauss law violation.
Figures~\ref{fig:GB_result_22} and~\ref{fig:GB_result_33} compare the resulting postselected gate-based data with the corresponding MBQS data.
The two implementations show qualitatively similar time dependence in the Gauss law generator, the 't Hooft operator, and the energy density.
In these data, the gate-based implementation shows a slower degradation over the accessible Trotter steps than the MBQS implementation.

\begin{figure}
	\centering
\includegraphics[width=0.9\linewidth]{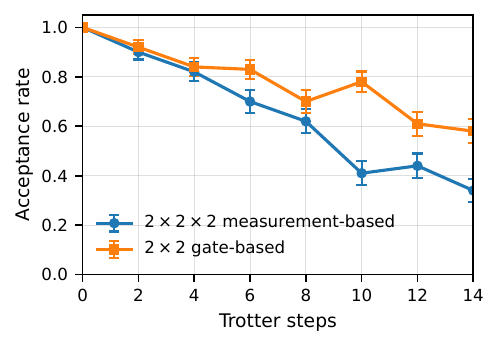}
	\caption{\textbf{Postselection acceptance rate for MBQS and gate-based simulation.}
	The acceptance rate $N_{\rm acc}/N_{\rm shot}$ is plotted versus the number of Trotter steps for the $(L_x,L_y)=(2,2)$ comparison.
	For MBQS, a shot is accepted only when all checked one-form-symmetry syndromes are trivial.
	For the gate-based simulation, a shot is accepted only when all measured Gauss-law syndromes are trivial.
	Each point uses $N_{\rm shot}=100$ raw shots.
	Error bars denote binomial standard errors, \(\sqrt{\widehat P_{\rm acc}(1-\widehat P_{\rm acc})/N_{\rm shot}}\).}
	\label{fig:acceptance_rate}
\end{figure}

Figure~\ref{fig:acceptance_rate} shows the postselection acceptance rate
$N_{\rm acc}/N_{\rm shot}$, where $N_{\rm acc}$ is the number of shots retained after applying the Gauss law syndrome criterion.
For both the gate-based simulation and MBQS, the acceptance rate decreases as the number of Trotter steps increases, reflecting the accumulation of detected syndrome violations.
The two protocols exhibit similar decay trends, with the gate-based simulation retaining a slightly larger fraction of shots over the range studied here.
Both implementations use projective mid-circuit measurements followed by postselection; neither applies active correction. They differ in the syndrome measured, i.e., $G({\bf x})$ directly in the gate-based circuit and $U(\mathcal S_v)$ in MBQS.
This gate-based implementation serves as a direct benchmark for the architecture demonstrated in this work.
As discussed in Sec.~\ref{sec:scale-resource} and in Sec.~\ref{sec:limit-outlook}, the advantage we attribute to MBQS is architectural---the model-specific resource state realizes the adaptive-measurement dynamics without compiling the Trotter circuit into universal-MBQC measurement patterns---while any end-to-end advantage over direct gate-based simulation is prospective and platform dependent.
No such on-device advantage is claimed from the present data.

\section{Quantifying the effect of syndrome postselection}
\label{sec:postselection-quantitative}

Here we quantify the agreement between a subset of the experimental data in 
Figure~\ref{fig:experiment_result}, before and after postselection, with the exact second-order Trotterized evolution at all displayed time points.
For a curve consisting of $K$ data points, let $r_i$ be the exact reference Trotter value and $\widehat O_i^{(s)}$ the experimental estimate, where $s\in\{\mathrm{raw},\mathrm{ps}\}$ denotes the raw or postselected data.
We define
\begin{align}
\operatorname{RMSE}_{s}
&=\left[\frac{1}{K}\sum_{i=1}^K \left(\widehat O_i^{(s)} - r_i \right)^2 \right]^{1/2} \,, \nonumber \\
\Delta_{\mathrm{RMSE}}
&=\operatorname{RMSE}_{\mathrm{raw}} - \operatorname{RMSE}_{\mathrm{ps}} \,.
\label{eq:postselection-rmse}
\end{align}

A positive $\Delta_{\mathrm{RMSE}}$ indicates better aggregate agreement after postselection.
We use the unweighted RMSE because it measures absolute displacement from the reference, whereas a statistic weighted by uncertainty ($\chi^2$) would measure agreement relative to sampling precision.
Related quantum-simulation experiments have assessed postselection using aggregate errors relative to exact references~\cite{rubin2025cqed,obrien2023purification} and bootstrap resampling~\cite{rubin2025cqed,chen2023fermions}.
Because our raw and postselected estimates share shots, we use paired resampling to obtain a confidence interval for the difference in their curve-level RMSEs.

We estimate finite-shot uncertainty by applying a paired bootstrap at the shot level~\cite{EfronTibshirani1994,Davison_Hinkley_1997}.
At each displayed time and, where applicable, for each measurement basis, the data comprise $N_{\rm shot}=100$ experimental shots.
The resampling unit is one shot record containing both the measurement bit string ($\bm{b}^{(a)}$ in Appendix~\ref{sec:obs-unc}) and the binary flag used to apply the relevant postselection criterion (for MBQS, $b_{\rm ps}$ in Section~\ref{sec:postselection}).
We generate $N_{\rm boot}=10^5$ bootstrap replicates.
A bootstrap replicate for a given curve is one synthetic realization of the entire analyzed data set used to construct that curve.
To generate one, independently at every displayed time and applicable measurement basis, we draw $N_{\rm shot}$ shot records with replacement from the corresponding observed data set, allowing the same record to be selected more than once.
From each replicate, we recompute the raw and postselected observable estimates at every displayed time.

For the energy curves plotted in panels (c) and (f) of Figs.~\ref{fig:GB_result_22} and~\ref{fig:GB_result_33}, Eqs.~\eqref{eq:Z2LGT_Hamiltonian}, \eqref{eq:avr_electric}, and~\eqref{eq:avr_plaquette} give
\begin{align}
-\frac{\langle H\rangle}{N_{\rm LGT}}
&=\overline X+\frac{\lambda}{2}\overline B
=\overline X+\overline B\,,
\qquad \lambda=2\,,
\label{eq:postselection-energy}
\end{align}
where $\overline X$ is the average electric contribution measured using $X$-basis shots and $\overline B=\overline{W_{1\times1}}$ is the average plaquette contribution measured using $Z$-basis shots.
Because the two measurement-basis data sets were acquired independently, they are resampled independently and then combined according to Eq.~\eqref{eq:postselection-energy}.
From the resulting bootstrap distribution, we report the central 95\% percentile interval of $\Delta_{\mathrm{RMSE}}$.
An interval entirely above (below) zero indicates a bootstrap-resolved increase (decrease) in agreement, whereas a comparison is inconclusive when its interval includes zero.

Table~\ref{tab:postselection-rmse} reports all non-diagnostic curves plotted in Fig.~\ref{fig:experiment_result} and in the energy panels of Figs.~\ref{fig:GB_result_22} and~\ref{fig:GB_result_33}.

\begin{table*}[t]
\centering
\begin{ruledtabular}
\begin{tabular}{lccccc}
Data set
& $N_{\rm acc}^{\min}$
& $\operatorname{RMSE}_{\rm raw}$
& $\operatorname{RMSE}_{\rm ps}$
& $\Delta_{\rm RMSE}$ [95\% interval]
& Closer \\
\hline
\multicolumn{6}{l}{\textit{Fig.~\ref{fig:experiment_result}: non-diagnostic
MBQS observables}} \\
(a) MBQS $(2,2,2)$, plaquette ($1\times1$ Wilson loop)
& 41 & 0.187 & 0.084 & 0.103 [0.052, 0.132]$^{*}$ & 5/7 \\
(b) MBQS $(3,3,1)$, Wilson loop $1\times1$
& 28 & 0.148 & 0.125 & 0.023 [$-0.012$, 0.050] & 6/8 \\
(b) MBQS $(3,3,1)$, Wilson loop $1\times2$ (both orientations)
& 28 & 0.183 & 0.155 & 0.028 [$-0.012$, 0.059] & 6/8 \\
(b) MBQS $(3,3,1)$, Wilson loop $2\times2$
& 28 & 0.206 & 0.168 & 0.039 [$-0.007$, 0.073] & 3/8 \\
(c) MBQS $(2,2,2)$, electric density $\overline X$
& 40 & 0.217 & 0.165 & 0.052 [0.010, 0.085]$^{*}$ & 3/7 \\
(d) MBQS $(3,3,1)$, electric density $\overline X$
& 19 & 0.170 & 0.114 & 0.057 [0.016, 0.085]$^{*}$ & 7/10 \\
\hline
\multicolumn{6}{l}{\textit{Fig.~\ref{fig:GB_result_22}: rescaled energy
$-\langle H\rangle/N_{\rm LGT}$}} \\
Gate-based $(2,2)$
& 55 & 0.217 & 0.108 & 0.109 [0.039, 0.152]$^{*}$ & 4/8 \\
MBQS $(2,2,2)$
& 34 & 0.327 & 0.210 & 0.116 [0.048, 0.163]$^{*}$ & 6/8 \\
\hline
\multicolumn{6}{l}{\textit{Fig.~\ref{fig:GB_result_33}: rescaled energy
$-\langle H\rangle/N_{\rm LGT}$}} \\
Gate-based $(3,3)$
& 28 & 0.176 & 0.144 & 0.032 [$-0.035$, 0.084] & 4/7 \\
MBQS $(3,3,1)$
& 39 & 0.153 & 0.131 & 0.022 [$-0.025$, 0.056] & 5/7 \\
\end{tabular}
\end{ruledtabular}
\caption{\textbf{Curve-level deviation from the exact second-order Trotterized reference.}
We summarize the deviation for all non-diagnostic curves plotted in Fig.~\ref{fig:experiment_result} and the energy panels of Figs.~\ref{fig:GB_result_22} and~\ref{fig:GB_result_33}, in which the reference appears as the black dashed curve.
$N_{\rm acc}^{\min}$ is the smallest accepted-shot count among the displayed points and, for an energy curve, among its two measurement-basis samples.
``Closer'' counts displayed points at which postselection gives a strictly smaller absolute deviation.
An asterisk marks a central 95\% bootstrap interval entirely above zero.
Rows that share experimental shots are correlated and should not be interpreted as independent tests.}
\label{tab:postselection-rmse}
\end{table*}

As shown in Table~\ref{tab:postselection-rmse}, the observed RMSE decreases for all ten non-diagnostic curves, and five increases in agreement are bootstrap-resolved.
This is not a uniform pointwise effect: postselection moves only a subset of the points closer to the reference, with ``Closer'' counts ranging from 3/8 to
6/8.
To check whether this conclusion depends on the choice of curve-deviation metric, we repeated the paired calculation using the mean absolute error from the reference defined as
\begin{equation}
\operatorname{MAE}_s=
\frac{1}{K}\sum_{i=1}^{K}
\left|\widehat O_i^{(s)}-r_i\right| \,.
\end{equation}
It also decreases for all ten curves, with three increases resolved under both metrics: the MBQS $(2,2,2)$ plaquette, the MBQS $(2,2,2)$ electric density, and the MBQS $(2,2,2)$ energy.
The increases for the MBQS $(3,3,1)$ electric-density curve and the gate-based $(2,2)$ energy curve are resolved by RMSE but inconclusive by mean absolute
error.
The remaining five comparisons are inconclusive under both metrics.
For every table row, each of the $10^5$ bootstrap replicates contained at least one accepted shot at every required time and measurement basis, so no replicate was discarded.

The smallest accepted-shot counts cited in Section~\ref{sec:limit-outlook}, $N_{\rm acc}=13$ and $11$, occur at the last two times of the $X$-basis diagnostic experiment in panels (b) and (c) of Fig.~\ref{fig:experiment_Gauss}.
By comparison, the smallest accepted-shot count among the ten curves in Table~\ref{tab:postselection-rmse} is $N_{\rm acc}=19$.
Applying the same paired analysis to the full eight-point Gauss-law and 't Hooft-loop diagnostic curves gives $\Delta_{\rm RMSE}=0.331$ with interval $[0.295,0.355]$ and $\Delta_{\rm RMSE}=0.276$ with interval $[0.246,0.298]$, respectively.
For the Gauss-law and 't Hooft-loop diagnostics, respectively, 2 and 3 of the $10^5$ bootstrap replicates contained no accepted shot at one or more displayed time points.
The postselected curve estimator is undefined for those replicates, which were omitted.
The empirical bootstrap is useful for these small accepted samples because it does not rely on a Gaussian approximation.

Empirical resampling cannot, however, create information beyond the observed records.
The resulting intervals therefore quantify finite-shot variation conditional on the measured empirical distributions; they do not include run-to-run device drift or other systematic experimental uncertainties.
The data sets are not independent because some hardware jobs contribute to more than one curve.
The ten rows should therefore not be interpreted as ten independent statistical tests.
The intervals are curve-by-curve uncertainty summaries and are not adjusted for multiple comparisons.
The supported conclusion is therefore that postselection lowers the observed aggregate deviation for all ten non-diagnostic curves, with bootstrap-resolved RMSE evidence for five and metric-robust, bootstrap-resolved evidence for three, rather than that it improves every point or every curve with equal statistical strength.

The statistics above are curve-level summaries of the benefit of postselection.
To support the coherence-window statement in Section~\ref{sec:postselection}, we apply the same shot-level bootstrap to a pointwise-in-time statistic of the postselected data alone.
For each curve in Fig.~\ref{fig:experiment_result} and each displayed time $T$, the running maximum deviation is $M(T)=\max_{t_i\le T}|\widehat O^{\rm ps}_i-r_i|$.
Using the same $10^5$ replicates---resampling every displayed time independently and applying postselection within each replicate---we evaluate $M(T)$ replicate by replicate, and $\Delta_{68}(T)$ of Eq.~\eqref{eq:coherence-window} is the 68th percentile of the resulting distribution, a one-sided $68\%$ upper confidence bound on all deviations up to $T$.
Because $M(T)$ is non-decreasing in $T$ for every replicate, $\Delta_{68}(T)$ is non-decreasing, so the times satisfying $\Delta_{68}(T)<\Delta_{\rm thr}$ form a prefix of the displayed times, ending at the window $T^*$ reported in Table~\ref{tab:coherence-window}.
The windows quoted in Section~\ref{sec:postselection} are the minima of $T^*$ over the observables of each experiment, and $T^*$ can only take displayed values.
Like every interval in this appendix, $\Delta_{68}(T)$ quantifies finite-shot variation conditional on the observed empirical distributions and does not include device drift or other systematic uncertainties.

\begin{table}[t]
\centering
\begin{ruledtabular}
\begin{tabular}{lccc}
Data set
& $\Delta_{68}(T^*_{0.20})$
& $T^*_{0.20}$
& $T^*_{0.25}$ \\
\hline
(a) MBQS $(2,2,2)$, plaquette & 0.198 & 1.44\footnotemark[1] & 1.44\footnotemark[1] \\
(b) MBQS $(3,3,1)$, Wilson loop $1\times1$ & 0.125 & 0.60 & 0.60 \\
(b) MBQS $(3,3,1)$, Wilson loop $1\times2$ & 0.153 & 0.60 & 0.60 \\
(b) MBQS $(3,3,1)$, Wilson loop $2\times2$ & 0.183 & 0.60 & 0.60 \\
(c) MBQS $(2,2,2)$, electric density $\overline X$ & 0.150 & 0.96 & 0.96 \\
(d) MBQS $(3,3,1)$, electric density $\overline X$ & 0.129 & 0.84 & 0.96 \\
\end{tabular}
\end{ruledtabular}
\footnotetext[1]{The threshold is not reached within the displayed window.}
\caption{\textbf{Coherence windows from the running maximum deviation.}
For each postselected curve in Fig.~\ref{fig:experiment_result}, $T^*_{\Delta_{\rm thr}}$ is the largest displayed time $T$ with $\Delta_{68}(T')<\Delta_{\rm thr}$ for every displayed $T'\le T$, for $\Delta_{\rm thr}=0.20$ and $0.25$; $\Delta_{68}(T^*_{0.20})$ is the bound attained at the $0.20$ window.
Curves sharing experimental shots are correlated and are not independent tests.}
\label{tab:coherence-window}
\end{table}

\section{Noisy-emulator benchmark of an MBQS protocol for the (1+1)D Ising model}
We investigate how device noise affects the performance of MBQS protocols by classically simulating noisy quantum circuits with the Quantinuum H2-2 emulator. The emulator noise parameters are calibrated to the component-level error rates measured on the H2-2 device, summarized in Table~\ref{tab:quantinuum_metrics}. The emulator also allows selected noise sources to be included while others are turned off, providing a controlled way to assess their relative impact.
However, a component-resolved noise study of the smallest nontrivial $(2+1)$D LGT instance would require repeating computationally demanding 24-qubit adaptive statevector simulations across multiple noise configurations.
We therefore use the $(1+1)$D transverse-field Ising model as a compact benchmark.

We consider a periodic chain of
$L_x$
 spins governed by the $(1+1)$D transverse-field Ising Hamiltonian
\begin{align}
H=-\lambda \sum_{j=1}^{L_x} Z_jZ_{j+1} -\sum_{j=1}^{L_x} X_j ,
\end{align}
with $Z_{L_x+1} \equiv Z_1$. Starting from $|\psi_{\rm Ising}(0)\rangle= |+\rangle^{\otimes L_x}
$, we define the simulated dynamics by the first-order Trotterized state
\begin{align}
|\psi_{\rm Ising}(t)\rangle
\equiv
\Bigg(
\prod_{j=1}^{L_x} e^{i \delta t X_j}
\prod_{j=1}^{L_x} e^{i\lambda \delta t Z_jZ_{j+1}}
\Bigg)^{N_t}
|\psi_{\rm Ising}(0)\rangle ,
\label{eq:Ising-Hamiltonian}
\end{align}
where $t=N_t\delta t$.

The first-order Trotterized dynamics defined in Eq.~\eqref{eq:Ising-Hamiltonian} can be implemented by adaptive measurements of a 2D cluster state. Consider a 2D square lattice with a periodic boundary condition in the $x$ direction and an open boundary condition in the $z$ direction, where $0\leq z\leq N_t$. We place qubits on the vertices $V$ and edges $E$ of this lattice. The input state is placed on the boundary vertices at $z=0$, while the remaining resource qubits are initialized in $|+\rangle$ on the set $\mathcal R$, consisting of edge qubits in the slabs $0\leq z<N_t$ and vertex qubits with $0<z\leq N_t$. The cluster state is
\begin{align}
|\Psi_{\rm 2D}\rangle
=
\prod_{e\in E}
\Bigl(
\prod_{v\subset e} {\rm CZ}_{e,v}
\Bigr)
|\psi_{\rm Ising}(0)\rangle
\otimes |+\rangle^{{\mathcal R}} \, .
\end{align}
The adaptive measurement pattern proceeds layer by layer. At layer $z$, measuring the $x$-directional edge qubits in the eigenbasis of $\mathcal{O}_A(\xi_1)=e^{-i\xi_1 X}Ze^{i\xi_1 X}$ implements the $Z_jZ_{j+1}$ part of the Trotter step, with $\xi_1=\pm\lambda\delta t$. The vertex qubits in the same layer are then measured in the $X$ basis, teleporting the state to the next layer and updating the Pauli byproducts. Measuring the $z$-directional edge qubits between layers $z$ and $z+1$ in the eigenbasis of $\mathcal{O}_B(\xi_2)=e^{-i\xi_2 Z}Xe^{i\xi_2 Z}$ implements the transverse-field part, with $\xi_2=\pm\delta t$. The signs of $\xi_1$ and $\xi_2$ are chosen adaptively from previous measurement outcomes. Repeating this pattern for $0\leq z<N_t$ produces the output state on the vertices at $z=N_t$, after which the tracked Pauli byproducts are removed by final Pauli corrections and the observable $O$ is measured.

\begin{figure}
	\centering
\includegraphics[width=\linewidth]{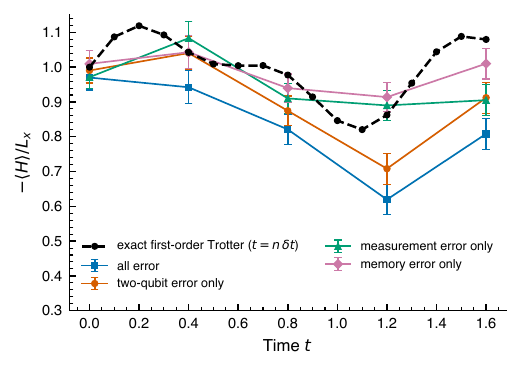}
	\caption{\textbf{Noisy H2-2 emulator benchmark for MBQS of the $(1+1)$D transverse-field Ising model.}
	We plot the rescaled energy expectation value $-\langle H\rangle/L_x$ for the first-order Trotterized dynamics defined in Eq.~\eqref{eq:Ising-Hamiltonian}, with $\lambda=1.6$, $\delta t=0.10$, initial state $|+\rangle^{\otimes 4}$, and $N_{\rm shot}=500$ shots.
	The MBQS protocol uses an instantaneous cluster-state block with $(L_x,L_z)=(4,1)$, containing 12 qubits.
	The exact first-order Trotter result is shown together with H2-2 emulator results obtained by selectively including all errors, only two-qubit gate errors, only measurement errors, or only memory errors.}
	\label{fig:Ising_result}
\end{figure}

Figure~\ref{fig:Ising_result} shows noisy classical-emulator results for this MBQS protocol on the Quantinuum H2-2 emulator. We use the $(L_x,L_z)=(4,1)$ instantaneous cluster-state block, which contains 12 qubits, and compare the exact first-order Trotterized evolution with emulator runs in which different noise sources are selectively included. The tuned-noise simulations show that measurement errors produce deviations comparable to those from two-qubit gate errors, rather than a parametrically larger degradation. This supports the use of mid-circuit measurements as a central computational primitive in MBQS protocols.

\bibliographystyle{apsrev4-2}
\bibliography{refs}

@article{Sivarajah:2020lfo,
    author = "Sivarajah, Seyon and Dilkes, Silas and Cowtan, Alexander and Simmons, Will and Edgington, Alec and Duncan, Ross",
    title = "{t|ket{\ensuremath{\rangle}}: a retargetable compiler for NISQ devices}",
    eprint = "2003.10611",
    archivePrefix = "arXiv",
    primaryClass = "quant-ph",
    doi = "10.1088/2058-9565/ab8e92",
    journal = "Quantum Sci. Technol.",
    volume = "6",
    number = "1",
    pages = "014003",
    year = "2020"
}

@ARTICLE{2026NatSR..16..689K,
       author = {{Kang}, Haiyue and {Kam}, John F. and {Mooney}, Gary J. and {Hollenberg}, Lloyd C.~L.},
        title = "{Entanglement teleportation along a regenerating hamster-wheel graph state}",
      journal = {Scientific Reports},
         year = 2026,
        month = dec,
       volume = {16},
          eid = {689},
        pages = {689},
          doi = {10.1038/s41598-025-30301-0},
       adsurl = {https://ui.adsabs.harvard.edu/abs/2026NatSR..16..689K}
}

@ARTICLE{2007PhRvL..98k7207D,
       author = {{den Nest}, M. Van and {D{\"u}r}, W. and {Briegel}, H.~J.},
        title = "{Classical Spin Models and the Quantum-Stabilizer Formalism}",
      journal = {\prl},
         year = 2007,
        month = mar,
       volume = {98},
       number = {11},
          eid = {117207},
        pages = {117207},
          doi = {10.1103/PhysRevLett.98.117207},
archivePrefix = {arXiv},
       eprint = {quant-ph/0610157},
 primaryClass = {quant-ph},
       adsurl = {https://ui.adsabs.harvard.edu/abs/2007PhRvL..98k7207D}
}

@article{rubin2025cqed,
  author       = {Rubin, Alex H. and Marinelli, Brian and Norman, Victoria A. and Rizvi, Zainab and Burch, Ashlyn Damaris and Naik, Ravi K. and Kreikebaum, John Mark and Chow, Matthew Nikolas Helson and Lobser, Daniel S. and Revelle, Melissa C. and others},
  title        = {Digital quantum simulation of cavity quantum electrodynamics: insights from superconducting and trapped ion quantum testbeds},
  doi          = {10.1088/2058-9565/ae0af0},
  url          = {https://www.osti.gov/biblio/2999151},
  journal      = {Quantum Science and Technology},
  issn         = {ISSN 2058-9565},
  number       = {4},
  volume       = {10},
  place        = {United States},
  publisher    = {IOP Publishing},
  year         = {2025},
  month        = {10}}

@book{Davison_Hinkley_1997, place={Cambridge}, series={Cambridge Series in Statistical and Probabilistic Mathematics}, title={Bootstrap Methods and their Application}, publisher={Cambridge University Press}, author={Davison, A. C. and Hinkley, D. V.}, year={1997}, collection={Cambridge Series in Statistical and Probabilistic Mathematics}}

@book{EfronTibshirani1994,
  author    = {Efron, Bradley and Tibshirani, Robert J.},
  title     = {An Introduction to the Bootstrap},
  edition   = {1},
  year      = {1994},
  publisher = {Chapman and Hall/CRC},
  doi       = {10.1201/9780429246593},
  url       = {https://doi.org/10.1201/9780429246593}
}

@ARTICLE{2002JMP....43.4452D,
       author = {{Dennis}, Eric and {Kitaev}, Alexei and {Landahl}, Andrew and {Preskill}, John},
        title = "{Topological quantum memory}",
      journal = {Journal of Mathematical Physics},
         year = 2002,
        month = sep,
       volume = {43},
       number = {9},
        pages = {4452-4505},
          doi = {10.1063/1.1499754},
archivePrefix = {arXiv},
       eprint = {quant-ph/0110143},
 primaryClass = {quant-ph},
       adsurl = {https://ui.adsabs.harvard.edu/abs/2002JMP....43.4452D}
}

@ARTICLE{Wegner,
       author = {{Wegner}, Franz J.},
        title = "{Duality in Generalized Ising Models and Phase Transitions without Local Order Parameters}",
      journal = {Journal of Mathematical Physics},
         year = 1971,
        month = oct,
       volume = {12},
       number = {10},
        pages = {2259-2272},
          doi = {10.1063/1.1665530},
       adsurl = {https://ui.adsabs.harvard.edu/abs/1971JMP....12.2259W}
}

@ARTICLE{2016Natur.534..516M,
       author = {{Martinez}, Esteban A. and {Muschik}, Christine A. and {Schindler}, Philipp and {Nigg}, Daniel and {Erhard}, Alexander and {Heyl}, Markus and {Hauke}, Philipp and {Dalmonte}, Marcello and {Monz}, Thomas and {Zoller}, Peter and {Blatt}, Rainer},
        title = "{Real-time dynamics of lattice gauge theories with a few-qubit quantum computer}",
      journal = {\nat},
         year = 2016,
        month = jun,
       volume = {534},
       number = {7608},
        pages = {516-519},
          doi = {10.1038/nature18318},
archivePrefix = {arXiv},
       eprint = {1605.04570},
 primaryClass = {quant-ph},
       adsurl = {https://ui.adsabs.harvard.edu/abs/2016Natur.534..516M}
}

@ARTICLE{2025NatCo..16.5492M,
       author = {{Mueller}, Niklas and {Wang}, Tianyi and {Katz}, Or and {Davoudi}, Zohreh and {Cetina}, Marko},
        title = "{Quantum computing universal thermalization dynamics in a (2 + 1)D Lattice Gauge Theory}",
      journal = {Nature Communications},
         year = 2025,
        month = jul,
       volume = {16},
          eid = {5492},
        pages = {5492},
          doi = {10.1038/s41467-025-60177-7},
archivePrefix = {arXiv},
       eprint = {2408.00069},
 primaryClass = {quant-ph}
}

@ARTICLE{2025arXiv250708088C,
       author = {{Cobos}, Jes{\'u}s and {Fraxanet}, Joana and {Benito}, C{\'e}sar and {di Marcantonio}, Francesco and {Rivero}, Pedro and {Kap{\'a}s}, Korn{\'e}l and {Werner}, Mikl{\'o}s Antal and {Legeza}, {\"O}rs and {Bermudez}, Alejandro and {Rico}, Enrique},
        title = "{Real-Time Dynamics in a (2+1)-D Gauge Theory: The Stringy Nature on a Superconducting Quantum Simulator}",
      journal = {arXiv e-prints},
         year = 2025,
        month = jul,
          eid = {arXiv:2507.08088},
        pages = {arXiv:2507.08088},
          doi = {10.48550/arXiv.2507.08088},
archivePrefix = {arXiv},
       eprint = {2507.08088},
 primaryClass = {quant-ph}
}

@ARTICLE{2025NatPh..21..570M,
       author = {{Meth}, Michael and {Zhang}, Jinglei and {Haase}, Jan F. and {Edmunds}, Claire and {Postler}, Lukas and {Jena}, Andrew J. and {Steiner}, Alex and {Dellantonio}, Luca and {Blatt}, Rainer and {Zoller}, Peter and {Monz}, Thomas and {Schindler}, Philipp and {Muschik}, Christine and {Ringbauer}, Martin},
        title = "{Simulating two-dimensional lattice gauge theories on a qudit quantum computer}",
      journal = {Nature Physics},
         year = 2025,
        month = mar,
       volume = {21},
       number = {4},
        pages = {570-576},
          doi = {10.1038/s41567-025-02797-w}
}

@misc{Aarts:2026uiu,
    author        = "Aarts, Gert and Sexty, D{\'e}nes",
    title         = "{Lattice field theories with a sign problem}",
    eprint        = "2604.24290",
    archivePrefix = "arXiv",
    primaryClass  = "hep-lat",
    year          = "2026",
    note          = "Chapter for the Encyclopedia of Nuclear Physics",
}

@article{Bauer:2022hpo,
    author = "Bauer, Christian W. and others",
    title = "{Quantum Simulation for High-Energy Physics}",
    eprint = "2204.03381",
    archivePrefix = "arXiv",
    primaryClass = "quant-ph",
    reportNumber = "UMD-PP-022-04, LA-UR-22-22100, RIKEN-iTHEMS-Report-22, RIKEN-iTHEMS-Report-22,
  FERMILAB-PUB-22-249-SQMS-T, FERMILAB-PUB-22-249-SQMS-T, IQuS@UW-21-027, MITRE-21-03848-2, FERMILAB-PUB-22-249-SQMS-T",
    doi = "10.1103/PRXQuantum.4.027001",
    journal = "PRX Quantum",
    volume = "4",
    number = "2",
    pages = "027001",
    year = "2023"
}

@article{FlavourLatticeAveragingGroupFLAG:2024oxs,
    author = "Aoki, Y. and others",
    collaboration = "Flavour Lattice Averaging Group (FLAG)",
    title = "{FLAG review 2024}",
    eprint = "2411.04268",
    archivePrefix = "arXiv",
    primaryClass = "hep-lat",
    reportNumber = "CERN-TH-2024-192, FERMILAB-PUB-24-0785-T",
    doi = "10.1103/nfzp-p5dn",
    journal = "Phys. Rev. D",
    volume = "113",
    number = "1",
    pages = "014508",
    year = "2026"
}

@article{Halimeh:2020gauge-reliability,
    author = "Halimeh, Jad C. and Hauke, Philipp",
    title = "{Reliability of lattice gauge theories}",
    eprint = "2001.00024",
    archivePrefix = "arXiv",
    primaryClass = "cond-mat.quant-gas",
    doi = "10.1103/PhysRevLett.125.030503",
    journal = "Phys. Rev. Lett.",
    volume = "125",
    pages = "030503",
    year = "2020"
}

@misc{QuantinuumPerformance2026,
  author = {{Quantinuum}},
  title = {Performance Validation},
  year = {2026},
  howpublished = {\url{https://docs.quantinuum.com/systems/user_guide/hardware_user_guide/performance_validation.html}},
  note = {{H2-2} hardware-specification data set dated 2025-08-28; accessed 2026-06-03}
}

@ARTICLE{2003PhRvA..68b2312R,
       author = {{Raussendorf}, Robert and {Browne}, Daniel E. and {Briegel}, Hans J.},
        title = "{Measurement-based quantum computation on cluster states}",
      journal = {\pra},
         year = 2003,
        month = aug,
       volume = {68},
       number = {2},
          eid = {022312},
        pages = {022312},
          doi = {10.1103/PhysRevA.68.022312},
archivePrefix = {arXiv},
       eprint = {quant-ph/0301052},
 primaryClass = {quant-ph},
       adsurl = {https://ui.adsabs.harvard.edu/abs/2003PhRvA..68b2312R}
}

@ARTICLE{2026arXiv260315561L,
       author = {{Lib}, Ohad and {Timme}, Hendrik and {Ammenwerth}, Maximilian and {Gyger}, Flavien and {Tao}, Renhao and {Sun}, Shijia and {Bloch}, Immanuel and {Zeiher}, Johannes},
        title = "{Velocity-Enabled Quantum Computing with Neutral Atoms}",
      journal = {arXiv e-prints},
         year = 2026,
        month = mar,
          eid = {arXiv:2603.15561},
        pages = {arXiv:2603.15561},
          doi = {10.48550/arXiv.2603.15561},
archivePrefix = {arXiv},
       eprint = {2603.15561},
 primaryClass = {quant-ph},
       adsurl = {https://ui.adsabs.harvard.edu/abs/2026arXiv260315561L}
}

@ARTICLE{2026PhRvL.136p0601Y,
       author = {{Yu}, Cheng-Cheng and {Chen}, Zi-Han and {Deng}, Yu-Hao and {Lu}, Chao-Yang and {Chen}, Ming-Cheng and {Pan}, Jian-Wei},
        title = "{Taming Rydberg Decay with Measurement-Based Quantum Computation}",
      journal = {\prl},
         year = 2026,
        month = apr,
       volume = {136},
       number = {16},
          eid = {160601},
        pages = {160601},
          doi = {10.1103/msbj-fxw7},
archivePrefix = {arXiv},
       eprint = {2411.04664},
 primaryClass = {quant-ph},
       adsurl = {https://ui.adsabs.harvard.edu/abs/2026PhRvL.136p0601Y}
}

@ARTICLE{2025arXiv250720009S,
       author = {{Stewart}, Luke M. and {Baranes}, Gefen and {Ramette}, Joshua and {Sinclair}, Josiah and {Vuleti{\'c}}, Vladan},
        title = "{Efficient construction of fault-tolerant neutral-atom cluster states}",
      journal = {arXiv e-prints},
         year = 2025,
        month = jul,
          eid = {arXiv:2507.20009},
        pages = {arXiv:2507.20009},
          doi = {10.48550/arXiv.2507.20009},
archivePrefix = {arXiv},
       eprint = {2507.20009},
 primaryClass = {quant-ph},
       adsurl = {https://ui.adsabs.harvard.edu/abs/2025arXiv250720009S}
}

@article{Edmonds1965PathsTA,
  title={Paths, Trees, and Flowers},
  author={Jack Edmonds},
  journal={Canadian Journal of Mathematics},
  year={1965},
  volume={17},
  pages={449 - 467},
  url={https://api.semanticscholar.org/CorpusID:18909734}
}

@ARTICLE{2015JHEP...02..172G,
       author = {{Gaiotto}, Davide and {Kapustin}, Anton and {Seiberg}, Nathan and {Willett}, Brian},
        title = "{Generalized global symmetries}",
      journal = {Journal of High Energy Physics},
         year = 2015,
        month = feb,
       volume = {2015},
          eid = {172},
        pages = {172},
          doi = {10.1007/JHEP02(2015)172},
archivePrefix = {arXiv},
       eprint = {1412.5148},
 primaryClass = {hep-th},
       adsurl = {https://ui.adsabs.harvard.edu/abs/2015JHEP...02..172G}
}

@ARTICLE{2016PhRvB..93o5131Y,
       author = {{Yoshida}, Beni},
        title = "{Topological phases with generalized global symmetries}",
      journal = {\prb},
         year = 2016,
        month = apr,
       volume = {93},
       number = {15},
          eid = {155131},
        pages = {155131},
          doi = {10.1103/PhysRevB.93.155131},
archivePrefix = {arXiv},
       eprint = {1508.03468},
 primaryClass = {cond-mat.str-el},
       adsurl = {https://ui.adsabs.harvard.edu/abs/2016PhRvB..93o5131Y}
}

@ARTICLE{2024PhRvX..14b1040T,
       author = {{Tantivasadakarn}, Nathanan and {Thorngren}, Ryan and {Vishwanath}, Ashvin and {Verresen}, Ruben},
        title = "{Long-Range Entanglement from Measuring Symmetry-Protected Topological Phases}",
      journal = {Physical Review X},
         year = 2024,
        month = jun,
       volume = {14},
       number = {2},
          eid = {021040},
        pages = {021040},
          doi = {10.1103/PhysRevX.14.021040},
archivePrefix = {arXiv},
       eprint = {2112.01519},
 primaryClass = {cond-mat.str-el},
       adsurl = {https://ui.adsabs.harvard.edu/abs/2024PhRvX..14b1040T}
}

@ARTICLE{2024CmPhy...7..205I,
       author = {{Iqbal}, Mohsin and {Tantivasadakarn}, Nathanan and {Gatterman}, Thomas M. and {Gerber}, Justin A. and {Gilmore}, Kevin and {Gresh}, Dan and {Hankin}, Aaron and {Hewitt}, Nathan and {Horst}, Chandler V. and {Matheny}, Mitchell and {Mengle}, Tanner and {Neyenhuis}, Brian and {Vishwanath}, Ashvin and {Foss-Feig}, Michael and {Verresen}, Ruben and {Dreyer}, Henrik},
        title = "{Topological order from measurements and feed-forward on a trapped ion quantum computer}",
      journal = {Communications Physics},
         year = 2024,
        month = dec,
       volume = {7},
       number = {1},
          eid = {205},
        pages = {205},
          doi = {10.1038/s42005-024-01698-3},
archivePrefix = {arXiv},
       eprint = {2302.01917},
 primaryClass = {quant-ph},
       adsurl = {https://ui.adsabs.harvard.edu/abs/2024CmPhy...7..205I}
}

@ARTICLE{2005PhRvA..71f2313R,
       author = {{Raussendorf}, Robert and {Bravyi}, Sergey and {Harrington}, Jim},
        title = "{Long-range quantum entanglement in noisy cluster states}",
      journal = {\pra},
         year = 2005,
        month = jun,
       volume = {71},
       number = {6},
          eid = {062313},
        pages = {062313},
          doi = {10.1103/PhysRevA.71.062313},
archivePrefix = {arXiv},
       eprint = {quant-ph/0407255},
 primaryClass = {quant-ph},
       adsurl = {https://ui.adsabs.harvard.edu/abs/2005PhRvA..71f2313R}
}

@ARTICLE{2022PhRvR...4c2013L,
       author = {{Lee}, Woo-Ram and {Qin}, Zhangjie and {Raussendorf}, Robert and {Sela}, Eran and {Scarola}, V.~W.},
        title = "{Measurement-based time evolution for quantum simulation of fermionic systems}",
      journal = {Physical Review Research},
         year = 2022,
        month = jul,
       volume = {4},
       number = {3},
          eid = {L032013},
        pages = {L032013},
          doi = {10.1103/PhysRevResearch.4.L032013},
archivePrefix = {arXiv},
       eprint = {2110.14642},
 primaryClass = {quant-ph},
       adsurl = {https://ui.adsabs.harvard.edu/abs/2022PhRvR...4c2013L}
}

@ARTICLE{2025NatCo..16..106R,
       author = {{Ringbauer}, Martin and {Hinsche}, Marcel and {Feldker}, Thomas and {Faehrmann}, Paul K. and {Bermejo-Vega}, Juani and {Edmunds}, Claire L. and {Postler}, Lukas and {Stricker}, Roman and {Marciniak}, Christian D. and {Meth}, Michael and {Pogorelov}, Ivan and {Blatt}, Rainer and {Schindler}, Philipp and {Eisert}, Jens and {Monz}, Thomas and {Hangleiter}, Dominik},
        title = "{Verifiable measurement-based quantum random sampling with trapped ions}",
      journal = {Nature Communications},
         year = 2025,
        month = jan,
       volume = {16},
       number = {1},
          eid = {106},
        pages = {106},
          doi = {10.1038/s41467-024-55342-3},
archivePrefix = {arXiv},
       eprint = {2307.14424},
 primaryClass = {quant-ph},
       adsurl = {https://ui.adsabs.harvard.edu/abs/2025NatCo..16..106R}
}

@ARTICLE{2005Natur.434..169W,
       author = {{Walther}, P. and {Resch}, K.~J. and {Rudolph}, T. and {Schenck}, E. and {Weinfurter}, H. and {Vedral}, V. and {Aspelmeyer}, M. and {Zeilinger}, A.},
        title = "{Experimental one-way quantum computing}",
      journal = {\nat},
         year = 2005,
        month = mar,
       volume = {434},
       number = {7030},
        pages = {169-176},
          doi = {10.1038/nature03347},
archivePrefix = {arXiv},
       eprint = {quant-ph/0503126},
 primaryClass = {quant-ph},
       adsurl = {https://ui.adsabs.harvard.edu/abs/2005Natur.434..169W}
}

@ARTICLE{2024PhRvL.132x0601C,
       author = {{Chan}, Albie and {Shi}, Zheng and {Dellantonio}, Luca and {D{\"u}r}, Wolfgang and {Muschik}, Christine A.},
        title = "{Measurement-Based Infused Circuits for Variational Quantum Eigensolvers}",
      journal = {\prl},
         year = 2024,
        month = jun,
       volume = {132},
       number = {24},
          eid = {240601},
        pages = {240601},
          doi = {10.1103/PhysRevLett.132.240601},
archivePrefix = {arXiv},
       eprint = {2305.19200},
 primaryClass = {quant-ph},
       adsurl = {https://ui.adsabs.harvard.edu/abs/2024PhRvL.132x0601C}
}

@ARTICLE{2024PhRvD.109k4508S,
       author = {{Schuster}, Stephan and {K{\"u}hn}, Stefan and {Funcke}, Lena and {Hartung}, Tobias and {Pleinert}, Marc-Oliver and {von Zanthier}, Joachim and {Jansen}, Karl},
        title = "{Studying the phase diagram of the three-flavor Schwinger model in the presence of a chemical potential with measurement- and gate-based quantum computing}",
      journal = {\prd},
         year = 2024,
        month = jun,
       volume = {109},
       number = {11},
          eid = {114508},
        pages = {114508},
          doi = {10.1103/PhysRevD.109.114508},
archivePrefix = {arXiv},
       eprint = {2311.14825},
 primaryClass = {hep-lat},
       adsurl = {https://ui.adsabs.harvard.edu/abs/2024PhRvD.109k4508S}
}

@ARTICLE{2013PhRvL.111u0501L,
       author = {{Lanyon}, B.~P. and {Jurcevic}, P. and {Zwerger}, M. and {Hempel}, C. and {Martinez}, E.~A. and {D{\"u}r}, W. and {Briegel}, H.~J. and {Blatt}, R. and {Roos}, C.~F.},
        title = "{Measurement-Based Quantum Computation with Trapped Ions}",
      journal = {\prl},
         year = 2013,
        month = nov,
       volume = {111},
       number = {21},
          eid = {210501},
        pages = {210501},
          doi = {10.1103/PhysRevLett.111.210501},
archivePrefix = {arXiv},
       eprint = {1308.5102},
 primaryClass = {quant-ph},
       adsurl = {https://ui.adsabs.harvard.edu/abs/2013PhRvL.111u0501L}
}

@ARTICLE{2025PhRvL.135p0801G,
       author = {{Gustiani}, Cica and {Leichtle}, Dominik and {Miller}, Jonathan and {Grassie}, Ross and {Mills}, Daniel and {Kashefi}, Elham},
        title = "{On-Chip Verified Quantum Computation with an Ion-Trap Quantum Processing Unit}",
      journal = {\prl},
         year = 2025,
        month = oct,
       volume = {135},
       number = {16},
          eid = {160801},
        pages = {160801},
          doi = {10.1103/jpms-v3kw},
archivePrefix = {arXiv},
       eprint = {2410.24133},
 primaryClass = {quant-ph},
       adsurl = {https://ui.adsabs.harvard.edu/abs/2025PhRvL.135p0801G}
}

@ARTICLE{2026NatPh..22..430J,
       author = {{Jiang}, Tao and {Cai}, Jianbin and {Huang}, Junxiang and {Zhou}, Naibin and {Zhang}, Yukun and {Bei}, Jiahao and {Cai}, Guoqing and {Cao}, Sirui and {Chen}, Fusheng and {Chen}, Jiang and {Chen}, Kefu and {Chen}, Xiawei and {Chen}, Xiqing and {Chen}, Zhe and {Chen}, Zhiyuan and {Chen}, Zihua and {Chu}, Wenhao and {Deng}, Hui and {Deng}, Zhibin and {Ding}, Pei and {Ding}, Xun and {Ding}, Zhuzhengqi and {Dong}, Shuai and {Fan}, Bo and {Fan}, Daojin and {Fu}, Yuanhao and {Gao}, Dongxin and {Ge}, Lei and {Gui}, Jiacheng and {Guo}, Cheng and {Guo}, Shaojun and {Guo}, Xiaoyang and {Han}, Lianchen and {He}, Tan and {Hong}, Linyin and {Hu}, Yisen and {Huang}, He-Liang and {Huo}, Yong-Heng and {Jiang}, Zuokai and {Jin}, Honghong and {Leng}, Yunxiang and {Li}, Dayu and {Li}, Dongdong and {Li}, Fangyu and {Li}, Jiaqi and {Li}, Jinjin and {Li}, Junyan and {Li}, Junyun and {Li}, Na and {Li}, Shaowei and {Li}, Wei and {Li}, Yuhuai and {Li}, Yuan and {Liang}, Futian and {Liang}, Xuelian and {Liao}, Nanxing and {Lin}, Jin and {Lin}, Weiping and {Liu}, Dailin and {Liu}, Hongxiu and {Liu}, Maliang and {Liu}, Xinyu and {Liu}, Xuemeng and {Liu}, Yancheng and {Lou}, Haoxin and {Ma}, Yuwei and {Meng}, Lingxin and {Mou}, Hao and {Nan}, Kailiang and {Nie}, Binghan and {Nie}, Meijuan and {Ning}, Jie and {Niu}, Le and {Peng}, Wenyi and {Qian}, Haoran and {Rong}, Hao and {Rong}, Tao and {Shen}, Huiyan and {Shen}, Qiong and {Su}, Hong and {Su}, Feifan and {Sun}, Chenyin and {Sun}, Liangchao and {Sun}, Tianzuo and {Sun}, Yingxiu and {Tan}, Yimeng and {Tan}, Jun and {Tang}, Longyue and {Tu}, Wenbing and {Wang}, Jiafei and {Wang}, Biao and {Wang}, Chang and {Wang}, Chen and {Wang}, Chu and {Wang}, Jian and {Wang}, Liangyuan and {Wang}, Rui and {Wang}, Shengtao and {Wang}, Xiaomin and {Wang}, Xinzhe and {Wang}, Xunxun and {Wang}, Yeru and {Wei}, Zuolin and {Wei}, Jiazhou and {Wu}, Dachao and {Wu}, Gang and {Wu}, Jin and {Wu}, Yulin and {Xie}, Shiyong and {Xin}, Lianjie and {Xu}, Yu and {Xue}, Chun and {Yan}, Kai and {Yang}, Weifeng and {Yang}, Xinpeng and {Yang}, Yang and {Ye}, Yangsen and {Ye}, Zhenping and {Ying}, Chong and {Yu}, Jiale and {Yu}, Qinjing and {Yu}, Wenhu and {Zeng}, Xiangdong and {Zha}, Chen and {Zhan}, Shaoyu and {Zhang}, Feifei and {Zhang}, Haibin and {Zhang}, Kaili and {Zhang}, Wen and {Zhang}, Yiming and {Zhang}, Yongzhuo and {Zhang}, Lixiang and {Zhao}, Guming and {Zhao}, Peng and {Zhao}, Xintao and {Zhao}, Youwei and {Zhao}, Zhong and {Zheng}, Luyuan and {Zhou}, Fei and {Zhou}, Liang and {Zhou}, Na and {Zhou}, Shifeng and {Zhou}, Shuang and {Zhou}, Zhengxiao and {Zhu}, Chengjun and {Zhu}, Qingling and {Zou}, Guihong and {Zou}, Haonan and {Zhang}, Qiang and {Lu}, Chao-Yang and {Peng}, Cheng-Zhi and {Yuan}, Xiao and {Gong}, Ming and {Zhu}, Xiaobo and {Pan}, Jian-Wei},
        title = "{One- and two-dimensional cluster states for topological phase simulation and measurement-based quantum computation}",
      journal = {Nature Physics},
         year = 2026,
        month = mar,
       volume = {22},
       number = {3},
        pages = {430-438},
          doi = {10.1038/s41567-026-03179-6},
archivePrefix = {arXiv},
       eprint = {2505.01978},
 primaryClass = {quant-ph},
       adsurl = {https://ui.adsabs.harvard.edu/abs/2026NatPh..22..430J}
}

@ARTICLE{2016RPPh...79a4401Z,
       author = {{Zohar}, Erez and {Cirac}, J. Ignacio and {Reznik}, Benni},
        title = "{Quantum simulations of lattice gauge theories using ultracold atoms in optical lattices}",
      journal = {Reports on Progress in Physics},
         year = 2016,
        month = jan,
       volume = {79},
       number = {1},
          eid = {014401},
        pages = {014401},
          doi = {10.1088/0034-4885/79/1/014401},
archivePrefix = {arXiv},
       eprint = {1503.02312},
 primaryClass = {quant-ph},
       adsurl = {https://ui.adsabs.harvard.edu/abs/2016RPPh...79a4401Z}
}

@book{Weinberg:1995mt,
    author = "Weinberg, Steven",
    title = "{The Quantum theory of fields. Vol. 1: Foundations}",
    doi = "10.1017/CBO9781139644167",
    isbn = "978-0-521-67053-1, 978-0-511-25204-4",
    publisher = "Cambridge University Press",
    month = "6",
    year = "2005"
}

@ARTICLE{2021Natur.592..209P,
       author = {{Pino}, J.~M. and {Dreiling}, J.~M. and {Figgatt}, C. and {Gaebler}, J.~P. and {Moses}, S.~A. and {Allman}, M.~S. and {Baldwin}, C.~H. and {Foss-Feig}, M. and {Hayes}, D. and {Mayer}, K. and {Ryan-Anderson}, C. and {Neyenhuis}, B.},
        title = "{Demonstration of the trapped-ion quantum CCD computer architecture}",
      journal = {\nat},
         year = 2021,
        month = apr,
       volume = {592},
       number = {7853},
        pages = {209-213},
          doi = {10.1038/s41586-021-03318-4},
archivePrefix = {arXiv},
       eprint = {2003.01293},
 primaryClass = {quant-ph},
       adsurl = {https://ui.adsabs.harvard.edu/abs/2021Natur.592..209P}
}

@ARTICLE{2001PhRvL..86..910B,
       author = {{Briegel}, Hans J. and {Raussendorf}, Robert},
        title = "{Persistent Entanglement in Arrays of Interacting Particles}",
      journal = {\prl},
         year = 2001,
        month = jan,
       volume = {86},
       number = {5},
        pages = {910-913},
          doi = {10.1103/PhysRevLett.86.910},
archivePrefix = {arXiv},
       eprint = {quant-ph/0004051},
 primaryClass = {quant-ph},
       adsurl = {https://ui.adsabs.harvard.edu/abs/2001PhRvL..86..910B}
}

@article{Bauer:2023qgm,
    author = "Bauer, Christian W. and Davoudi, Zohreh and Klco, Natalie and Savage, Martin J.",
    title = "{Quantum simulation of fundamental particles and forces}",
    eprint = "2404.06298",
    archivePrefix = "arXiv",
    primaryClass = "hep-ph",
    reportNumber = "IQuS@UW-21-052",
    doi = "10.1038/s42254-023-00599-8",
    journal = "Nature Rev. Phys.",
    volume = "5",
    number = "7",
    pages = "420--432",
    year = "2023"
}

@article{Banuls:2019bmf,
    author = "Ba{\~n}uls, M. C. and others",
    title = "{Simulating Lattice Gauge Theories within Quantum Technologies}",
    eprint = "1911.00003",
    archivePrefix = "arXiv",
    primaryClass = "quant-ph",
    doi = "10.1140/epjd/e2020-100571-8",
    journal = "Eur. Phys. J. D",
    volume = "74",
    number = "8",
    pages = "165",
    year = "2020"
}

@article{Kronfeld:2012uk,
    author = "Kronfeld, Andreas S.",
    title = "{Twenty-first Century Lattice Gauge Theory: Results from the QCD Lagrangian}",
    eprint = "1203.1204",
    archivePrefix = "arXiv",
    primaryClass = "hep-lat",
    reportNumber = "FERMILAB-PUB-12-064-T",
    doi = "10.1146/annurev-nucl-102711-094942",
    journal = "Ann. Rev. Nucl. Part. Sci.",
    volume = "62",
    pages = "265--284",
    year = "2012"
}

@book{Creutz:1983njd, place={Cambridge}, series={Cambridge Monographs on Mathematical Physics}, title={Quarks, Gluons and Lattices}, publisher={Cambridge University Press}, author={Creutz, Michael}, year={2023}, collection={Cambridge Monographs on Mathematical Physics}}

@article{Wilson:1974sk,
    author = "Wilson, Kenneth G.",
    editor = "Taylor, J. C.",
    title = "{Confinement of Quarks}",
    reportNumber = "CLNS-262",
    doi = "10.1103/PhysRevD.10.2445",
    journal = "Phys. Rev. D",
    volume = "10",
    pages = "2445--2459",
    year = "1974"
}

@ARTICLE{2026arXiv260113530H,
       author = {{Hayata}, Tomoya and {Hidaka}, Yoshimasa and {Kikuchi}, Yuta},
        title = "{Onset of thermalization of q-deformed SU(2) Yang-Mills theory on a trapped-ion quantum computer}",
      journal = {arXiv e-prints},
         year = 2026,
        month = jan,
          eid = {arXiv:2601.13530},
        pages = {arXiv:2601.13530},
          doi = {10.48550/arXiv.2601.13530},
archivePrefix = {arXiv},
       eprint = {2601.13530},
 primaryClass = {hep-lat},
       adsurl = {https://ui.adsabs.harvard.edu/abs/2026arXiv260113530H}
}

@ARTICLE{2023PhRvL.131q1902Z,
       author = {{Zache}, Torsten V. and {Gonz{\'a}lez-Cuadra}, Daniel and {Zoller}, Peter},
        title = "{Quantum and Classical Spin-Network Algorithms for q -Deformed Kogut-Susskind Gauge Theories}",
      journal = {\prl},
         year = 2023,
        month = oct,
       volume = {131},
       number = {17},
          eid = {171902},
        pages = {171902},
          doi = {10.1103/PhysRevLett.131.171902},
archivePrefix = {arXiv},
       eprint = {2304.02527},
 primaryClass = {quant-ph},
       adsurl = {https://ui.adsabs.harvard.edu/abs/2023PhRvL.131q1902Z}
}

@ARTICLE{2023JHEP...09..123H,
       author = {{Hayata}, Tomoya and {Hidaka}, Yoshimasa},
        title = "{q deformed formulation of Hamiltonian SU(3) Yang-Mills theory}",
      journal = {Journal of High Energy Physics},
         year = 2023,
        month = sep,
       volume = {2023},
       number = {9},
          eid = {123},
        pages = {123},
          doi = {10.1007/JHEP09(2023)123},
archivePrefix = {arXiv},
       eprint = {2306.12324},
 primaryClass = {hep-lat},
       adsurl = {https://ui.adsabs.harvard.edu/abs/2023JHEP...09..123H}
}

@ARTICLE{2023JHEP...09..126H,
       author = {{Hayata}, Tomoya and {Hidaka}, Yoshimasa},
        title = "{String-net formulation of Hamiltonian lattice Yang-Mills theories and quantum many-body scars in a nonabelian gauge theory}",
      journal = {Journal of High Energy Physics},
         year = 2023,
        month = sep,
       volume = {2023},
       number = {9},
          eid = {126},
        pages = {126},
          doi = {10.1007/JHEP09(2023)126},
archivePrefix = {arXiv},
       eprint = {2305.05950},
 primaryClass = {hep-lat},
       adsurl = {https://ui.adsabs.harvard.edu/abs/2023JHEP...09..126H}
}

@ARTICLE{2023NatCo..14.2242H,
       author = {{Hrmo}, Pavel and {Wilhelm}, Benjamin and {Gerster}, Lukas and {van Mourik}, Martin W. and {Huber}, Marcus and {Blatt}, Rainer and {Schindler}, Philipp and {Monz}, Thomas and {Ringbauer}, Martin},
        title = "{Native qudit entanglement in a trapped ion quantum processor}",
      journal = {Nature Communications},
         year = 2023,
        month = apr,
       volume = {14},
          eid = {2242},
        pages = {2242},
          doi = {10.1038/s41467-023-37375-2},
archivePrefix = {arXiv},
       eprint = {2206.04104},
 primaryClass = {quant-ph},
       adsurl = {https://ui.adsabs.harvard.edu/abs/2023NatCo..14.2242H}
}

@ARTICLE{2022NatPh..18.1053R,
       author = {{Ringbauer}, Martin and {Meth}, Michael and {Postler}, Lukas and {Stricker}, Roman and {Blatt}, Rainer and {Schindler}, Philipp and {Monz}, Thomas},
        title = "{A universal qudit quantum processor with trapped ions}",
      journal = {Nature Physics},
         year = 2022,
        month = sep,
       volume = {18},
       number = {9},
        pages = {1053-1057},
          doi = {10.1038/s41567-022-01658-0},
archivePrefix = {arXiv},
       eprint = {2109.06903},
 primaryClass = {quant-ph},
       adsurl = {https://ui.adsabs.harvard.edu/abs/2022NatPh..18.1053R}
}

@ARTICLE{2023ScPP...14..129S,
       author = {{Sukeno}, Hiroki and {Okuda}, Takuya},
        title = "{Measurement-based quantum simulation of Abelian lattice gauge theories}",
      journal = {SciPost Physics},
         year = 2023,
        month = may,
       volume = {14},
       number = {5},
          eid = {129},
        pages = {129},
          doi = {10.21468/SciPostPhys.14.5.129},
archivePrefix = {arXiv},
       eprint = {2210.10908},
 primaryClass = {quant-ph},
       adsurl = {https://ui.adsabs.harvard.edu/abs/2023ScPP...14..129S}
}

@ARTICLE{2024ScPP...17..113O,
       author = {{Okuda}, Takuya and {Parayil Mana}, Aswin and {Sukeno}, Hiroki},
        title = "{Anomaly inflow for CSS and fractonic lattice models and dualities via cluster state measurement}",
      journal = {SciPost Physics},
         year = 2024,
        month = oct,
       volume = {17},
       number = {4},
          eid = {113},
        pages = {113},
          doi = {10.21468/SciPostPhys.17.4.113},
archivePrefix = {arXiv},
       eprint = {2405.15853},
 primaryClass = {quant-ph},
       adsurl = {https://ui.adsabs.harvard.edu/abs/2024ScPP...17..113O}
}

@ARTICLE{2024PhRvR...6d3018O,
       author = {{Okuda}, Takuya and {Parayil Mana}, Aswin and {Sukeno}, Hiroki},
        title = "{Anomaly inflow, dualities, and quantum simulation of Abelian lattice gauge theories induced by measurements}",
      journal = {Physical Review Research},
         year = 2024,
        month = oct,
       volume = {6},
       number = {4},
          eid = {043018},
        pages = {043018},
          doi = {10.1103/PhysRevResearch.6.043018},
archivePrefix = {arXiv},
       eprint = {2402.08720},
 primaryClass = {cond-mat.str-el},
       adsurl = {https://ui.adsabs.harvard.edu/abs/2024PhRvR...6d3018O}
}

@article{RevModPhys.51.659,
  title = {An introduction to lattice gauge theory and spin systems},
  author = {Kogut, John B.},
  journal = {Rev. Mod. Phys.},
  volume = {51},
  issue = {4},
  pages = {659--713},
  numpages = {0},
  year = {1979},
  month = {Oct},
  publisher = {American Physical Society},
  doi = {10.1103/RevModPhys.51.659},
  url = {https://link.aps.org/doi/10.1103/RevModPhys.51.659}
}

@article{DeCross:2022kuu,
    author = "DeCross, Matthew and Chertkov, Eli and Kohagen, Megan and Foss-Feig, Michael",
    title = "{Qubit-Reuse Compilation with Mid-Circuit Measurement and Reset}",
    eprint = "2210.08039",
    archivePrefix = "arXiv",
    primaryClass = "quant-ph",
    doi = "10.1103/PhysRevX.13.041057",
    journal = "Phys. Rev. X",
    volume = "13",
    number = "4",
    pages = "041057",
    year = "2023"
}

@article{Wineland:1997mg,
    author = "Wineland, D. J. and Monroe, C. and Itano, W. M. and Leibfried, D. and King, B. E. and Meekhof, D. M.",
    title = "{Experimental issues in coherent quantum-state manipulation of trapped atomic ions}",
    eprint = "quant-ph/9710025",
    archivePrefix = "arXiv",
    doi = "10.6028/jres.103.019",
    journal = "J. Res. Natl. Inst. Stand. Tech.",
    volume = "103",
    number = "3",
    pages = "259",
    year = "1998"
}

@article{Kielpinski:2002wbd,
    author = "Kielpinski, D. and Monroe, C. and Wineland, D. J.",
    title = "{Architecture for a large-scale ion-trap quantum computer}",
    doi = "10.1038/nature00784",
    journal = "Nature",
    volume = "417",
    pages = "709--711",
    year = "2002"
}

@article{PhysRevX.13.041052,
  title = {A Race-Track Trapped-Ion Quantum Processor},
  author = {Moses, S. A. and and Baldwin, C. H. and Allman, M. S. and Ancona, R. and Ascarrunz, L. and Barnes, C. and Bartolotta, J. and Bjork, B. and Blanchard, P. and Bohn, M. and Bohnet, J. G. and Brown, N. C. and Burdick, N. Q. and Burton, W. C. and Campbell, S. L. and Campora, J. P. and Carron, C. and Chambers, J. and Chan, J. W. and Chen, Y. H. and Chernoguzov, A. and Chertkov, E. and Colina, J. and Curtis, J. P. and Daniel, R. and DeCross, M. and Deen, D. and Delaney, C. and Dreiling, J. M. and Ertsgaard, C. T. and Esposito, J. and Estey, B. and Fabrikant, M. and Figgatt, C. and Foltz, C. and Foss-Feig, M. and Francois, D. and Gaebler, J. P. and Gatterman, T. M. and Gilbreth, C. N. and Giles, J. and Glynn, E. and Hall, A. and Hankin, A. M. and Hansen, A. and Hayes, D. and Higashi, B. and Hoffman, I. M. and Horning, B. and Hout, J. J. and Jacobs, R. and Johansen, J. and Jones, L. and Karcz, J. and Klein, T. and Lauria, P. and Lee, P. and Liefer, D. and Lu, S. T. and Lucchetti, D. and Lytle, C. and Malm, A. and Matheny, M. and Mathewson, B. and Mayer, K. and Miller, D. B. and Mills, M. and Neyenhuis, B. and Nugent, L. and Olson, S. and Parks, J. and Price, G. N. and Price, Z. and Pugh, M. and Ransford, A. and Reed, A. P. and Roman, C. and Rowe, M. and Ryan-Anderson, C. and Sanders, S. and Sedlacek, J. and Shevchuk, P. and Siegfried, P. and Skripka, T. and Spaun, B. and Sprenkle, R. T. and Stutz, R. P. and Swallows, M. and Tobey, R. I. and Tran, A. and Tran, T. and Vogt, E. and Volin, C. and Walker, J. and Zolot, A. M. and Pino, J. M.},
  journal = {Phys. Rev. X},
  volume = {13},
  issue = {4},
  pages = {041052},
  numpages = {25},
  year = {2023},
  month = {Dec},
  publisher = {American Physical Society},
  doi = {10.1103/PhysRevX.13.041052},
  url = {https://link.aps.org/doi/10.1103/PhysRevX.13.041052}
}

@book{NielsenChuang,
    author = "Nielsen, Michael A. and Chuang, Isaac L.",
    title = "{Quantum Computation and Quantum Information: 10th Anniversary Edition}",
    publisher = "Cambridge University Press",
    year = "2010",
    doi = "10.1017/CBO9780511976667",
    isbn = "9781107002173",
    url = "https://doi.org/10.1017/CBO9780511976667"
}

@Article{obrien2023purification,
author="O'Brien, T. E.
and Anselmetti, G.
and Gkritsis, F.
and Elfving, V. E.
and Polla, S.
and Huggins, W. J.
and Oumarou, O.
and Kechedzhi, K.
and Abanin, D.
and Acharya, R.
and Aleiner, I.
and Allen, R.
and Andersen, T. I.
and Anderson, K.
and Ansmann, M.
and Arute, F.
and Arya, K.
and Asfaw, A.
and Atalaya, J.
and Bardin, J. C.
and Bengtsson, A.
and Bortoli, G.
and Bourassa, A.
and Bovaird, J.
and Brill, L.
and Broughton, M.
and Buckley, B.
and Buell, D. A.
and Burger, T.
and Burkett, B.
and Bushnell, N.
and Campero, J.
and Chen, Z.
and Chiaro, B.
and Chik, D.
and Cogan, J.
and Collins, R.
and Conner, P.
and Courtney, W.
and Crook, A. L.
and Curtin, B.
and Debroy, D. M.
and Demura, S.
and Drozdov, I.
and Dunsworth, A.
and Erickson, C.
and Faoro, L.
and Farhi, E.
and Fatemi, R.
and Ferreira, V. S.
and Flores Burgos, L.
and Forati, E.
and Fowler, A. G.
and Foxen, B.
and Giang, W.
and Gidney, C.
and Gilboa, D.
and Giustina, M.
and Gosula, R.
and Grajales Dau, A.
and Gross, J. A.
and Habegger, S.
and Hamilton, M. C.
and Hansen, M.
and Harrigan, M. P.
and Harrington, S. D.
and Heu, P.
and Hoffmann, M. R.
and Hong, S.
and Huang, T.
and Huff, A.
and Ioffe, L. B.
and Isakov, S. V.
and Iveland, J.
and Jeffrey, E.
and Jiang, Z.
and Jones, C.
and Juhas, P.
and Kafri, D.
and Khattar, T.
and Khezri, M.
and Kieferov{\'a}, M.
and Kim, S.
and Klimov, P. V.
and Klots, A. R.
and Korotkov, A. N.
and Kostritsa, F.
and Kreikebaum, J. M.
and Landhuis, D.
and Laptev, P.
and Lau, K.-M.
and Laws, L.
and Lee, J.
and Lee, K.
and Lester, B. J.
and Lill, A. T.
and Liu, W.
and Livingston, W. P.
and Locharla, A.
and Malone, F. D.
and Mandr{\`a}, S.
and Martin, O.
and Martin, S.
and McClean, J. R.
and McCourt, T.
and McEwen, M.
and Mi, X.
and Mieszala, A.
and Miao, K. C.
and Mohseni, M.
and Montazeri, S.
and Morvan, A.
and Movassagh, R.
and Mruczkiewicz, W.
and Naaman, O.
and Neeley, M.
and Neill, C.
and Nersisyan, A.
and Newman, M.
and Ng, J. H.
and Nguyen, A.
and Nguyen, M.
and Niu, M. Y.
and Omonije, S.
and Opremcak, A.
and Petukhov, A.
and Potter, R.
and Pryadko, L. P.
and Quintana, C.
and Rocque, C.
and Roushan, P.
and Saei, N.
and Sank, D.
and Sankaragomathi, K.
and Satzinger, K. J.
and Schurkus, H. F.
and Schuster, C.
and Shearn, M. J.
and Shorter, A.
and Shutty, N.
and Shvarts, V.
and Skruzny, J.
and Smith, W. C.
and Somma, R. D.
and Sterling, G.
and Strain, D.
and Szalay, M.
and Thor, D.
and Torres, A.
and Vidal, G.
and Villalonga, B.
and Vollgraff Heidweiller, C.
and White, T.
and Woo, B. W. K.
and Xing, C.
and Yao, Z. J.
and Yeh, P.
and Yoo, J.
and Young, G.
and Zalcman, A.
and Zhang, Y.
and Zhu, N.
and Zobrist, N.
and Bacon, D.
and Boixo, S.
and Chen, Y.
and Hilton, J.
and Kelly, J.
and Lucero, E.
and Megrant, A.
and Neven, H.
and Smelyanskiy, V.
and Gogolin, C.
and Babbush, R.
and Rubin, N. C.",
title="Purification-based quantum error mitigation of pair-correlated electron simulations",
journal="Nature Physics",
year="2023",
month="Dec",
day="01",
volume="19",
number="12",
pages="1787--1792",
issn="1745-2481",
doi="10.1038/s41567-023-02240-y",
url="https://doi.org/10.1038/s41567-023-02240-y"
}

@Article{chen2023fermions,
author="Chen, Wentao
and Zhang, Shuaining
and Zhang, Jialiang
and Su, Xiaolu
and Lu, Yao
and Zhang, Kuan
and Qiao, Mu
and Li, Ying
and Zhang, Jing-Ning
and Kim, Kihwan",
title="Error-mitigated quantum simulation of interacting fermions with trapped ions",
journal="npj Quantum Information",
year="2023",
month="Dec",
day="07",
volume="9",
number="1",
pages="122",
issn="2056-6387",
doi="10.1038/s41534-023-00784-8",
url="https://doi.org/10.1038/s41534-023-00784-8"
}
\end{document}